\documentclass[twocolumn, tighten, times, astrosymb]{aastex631}

\usepackage{xcolor}

\newcommand{\Teff}{\ensuremath{T_{\text{eff}}}}

\newcommand\dracoii{Draco~{\sc II}}
\newcommand\jwst{\textit{JWST}}
\newcommand\hst{\textit{HST}}
\newcommand\msun{$M_{\odot}$}
\newcommand\mathmsun{M_{\odot}}

\shorttitle{JWST Resolved Stellar Populations I}
\shortauthors{Weisz et al.}
\graphicspath{{./}{figures/}}

\begin{document}

\title{The JWST Resolved Stellar Populations Early Release Science Program II. \\ Survey Overview}

\correspondingauthor{Daniel R. Weisz}
\email{dan.weisz@berkeley.edu}

\author[0000-0002-6442-6030]{Daniel R. Weisz}
\affiliation{Department of Astronomy, University of California, Berkeley, CA 94720, USA}

\author[0000-0001-5538-2614]{Kristen B. W. McQuinn}
\affiliation{Department of Physics and Astronomy, Rutgers, the State University of New Jersey,  136 Frelinghuysen Road, Piscataway, NJ 08854, USA}

\author[0000-0002-1445-4877]{Alessandro Savino}
\affiliation{Department of Astronomy, University of California, Berkeley, CA 94720, USA}

\author[0000-0002-3204-1742]{Nitya Kallivayalil}
\affiliation{Department of Astronomy, University of Virginia, 530 McCormick Road, Charlottesville, VA 22904, USA}

\author[0000-0003-2861-3995]{Jay Anderson}
\affiliation{Space Telescope Science Institute, 3700 San Martin Drive, Baltimore, MD 21218, USA}

\author[0000-0003-4850-9589]{Martha L. Boyer}
\affiliation{Space Telescope Science Institute, 3700 San Martin Drive, Baltimore, MD 21218, USA}

\author[0000-0001-6464-3257]{Matteo Correnti}
\affiliation{INAF Osservatorio Astronomico di Roma, Via Frascati 33, 00078, Monteporzio Catone, Rome, Italy}
\affiliation{ASI-Space Science Data Center, Via del Politecnico, I-00133, Rome, Italy}

\author[0000-0002-7007-9725]{Marla C. Geha}
\affiliation{Department of Astronomy, Yale University, New Haven, CT 06520, USA}

\author[0000-0001-8416-4093]{Andrew E. Dolphin}
\affiliation{Raytheon Technologies, 1151 E. Hermans Road, Tucson, AZ 85756, USA}
\affiliation{Steward Observatory, University of Arizona, 933 N. Cherry Avenue, Tucson, AZ 85719, USA}

\author[0000-0002-4378-8534]{Karin M. Sandstrom}
\affiliation{Center for Astrophysics and Space Sciences, Department of Physics, University of California San Diego, 9500 Gilman Drive, La Jolla, CA 92093, USA}

\author[0000-0003-0303-3855]{Andrew A. Cole}
\affiliation{School of Natural Sciences, University of Tasmania, Private Bag 37, Hobart, Tasmania 7001, Australia}

\author[0000-0002-7502-0597]{Benjamin F. Williams}
\affiliation{Department of Astronomy, University of Washington, Box 351580, U.W., Seattle, WA 98195-1580, USA}

\author[0000-0003-0605-8732]{Evan D. Skillman}
\affiliation{University of Minnesota, Minnesota Institute for Astrophysics, School of Physics and Astronomy, 116 Church Street, S.E., Minneapolis,
MN 55455, USA}

\author[0000-0002-2970-7435]{Roger E. Cohen}
\affiliation{Department of Physics and Astronomy, Rutgers, the State University of New Jersey,  136 Frelinghuysen Road, Piscataway, NJ 08854, USA}

\author[0000-0002-8092-2077]{Max J. B. Newman}
\affiliation{Department of Physics and Astronomy, Rutgers, the State University of New Jersey,  136 Frelinghuysen Road, Piscataway, NJ 08854, USA}

\author[0000-0002-1691-8217]{Rachael Beaton}
\affiliation{Space Telescope Science Institute, 3700 San Martin Drive, Baltimore, MD 21218, USA}
\affiliation{Department of Astrophysical Sciences, Princeton University, 4 Ivy Lane, Princeton, NJ 08544, USA}
\affiliation{The Observatories of the Carnegie Institution for Science, 813 Santa Barbara St., Pasadena, CA 91101, USA}

\author[0000-0002-7922-8440]{Alessandro Bressan}
\affiliation{SISSA, Via Bonomea 265, 34136 Trieste, Italy}

\author[0000-0002-5480-5686]{Alberto Bolatto}
\affiliation{Department of Astronomy, University of Maryland, College Park, MD 20742, USA}
\affiliation{Joint Space-Science Institute, University of Maryland, College Park, MD 20742, USA}

\author[0000-0002-9604-343X]{Michael Boylan-Kolchin}
\affiliation{Department of Astronomy, The University of Texas at Austin, 2515 Speedway, Stop C1400, Austin, TX 78712-1205, USA}

\author[0000-0002-0372-3736]{Alyson M. Brooks}
\affiliation{Department of Physics and Astronomy, Rutgers, the State University of New Jersey,  136 Frelinghuysen Road, Piscataway, NJ 08854, USA}
\affiliation{Center for Computational Astrophysics, Flatiron Institute, 162 Fifth Avenue, New York, NY 10010, USA}

\author[0000-0003-4298-5082]{James S. Bullock}
\affiliation{Department of Physics and Astronomy, University of California, Irvine, CA 92697 USA}

\author[0000-0002-1590-8551]{Charlie Conroy}
\affiliation{Center for Astrophysics | Harvard \& Smithsonian, Cambridge, MA, 02138, USA}

\author[0000-0003-1371-6019]{M. C. Cooper}
\affiliation{Department of Physics and Astronomy, University of California, Irvine, CA 92697 USA}

\author[0000-0002-1264-2006]{Julianne J. Dalcanton}
\affiliation{Department of Astronomy, University of Washington, Box 351580, U.W., Seattle, WA 98195-1580, USA}
\affiliation{Center for Computational Astrophysics, Flatiron Institute, 162 Fifth Avenue, New York, NY 10010, USA}

\author[0000-0002-4442-5700]{Aaron L. Dotter}
\affiliation{Department of Physics and Astronomy, Dartmouth College, 6127 Wilder Laboratory, Hanover, NH 03755, USA}

\author[0000-0002-3122-300X]{Tobias K. Fritz}
\affiliation{Department of Astronomy, University of Virginia, 530 McCormick Road, Charlottesville, VA 22904, USA}

\author[0000-0001-9061-1697]{Christopher T. Garling}
\affiliation{Department of Astronomy, University of Virginia, 530 McCormick Road, Charlottesville, VA 22904, USA}

\author[0000-0002-5581-2896]{Mario Gennaro}
\affiliation{Space Telescope Science Institute, 3700 San Martin Dr., Baltimore, MD 21218, USA}
\affiliation{The William H. Miller {\sc III} Department of Physics \& Astronomy, Bloomberg Center for Physics and Astronomy, Johns Hopkins University, 3400 N. Charles Street, Baltimore, MD 21218, USA}

\author[0000-0003-0394-8377]{Karoline M. Gilbert}
\affiliation{Space Telescope Science Institute, 3700 San Martin Dr., Baltimore, MD 21218, USA}
\affiliation{The William H. Miller {\sc III} Department of Physics \& Astronomy, Bloomberg Center for Physics and Astronomy, Johns Hopkins University, 3400 N. Charles Street, Baltimore, MD 21218, USA}

\author[0000-0002-6301-3269]{L\'eo Girardi}
\affiliation{Padova Astronomical Observatory, Vicolo dell'Osservatorio 5, Padova, Italy}

\author[0000-0002-9280-7594]{Benjamin D. Johnson}
\affiliation{Center for Astrophysics | Harvard \& Smithsonian, Cambridge, MA, 02138, USA}

\author[0000-0001-6421-0953]{L. Clifton Johnson}
\affiliation{Center for Interdisciplinary Exploration and Research in Astrophysics (CIERA) and Department of Physics and Astronomy, Northwestern University, 1800 Sherman Avenue, Evanston, IL 60201, USA}

\author[0000-0001-9690-4159]{Jason S. Kalirai}
\affiliation{John Hopkins Applied Physics Laboratory, 11100 Johns Hopkins Road, Laurel, MD 20723, USA}

\author[0000-0001-6196-5162]{Evan N. Kirby}
\affiliation{Department of Physics, University of Notre Dame, Notre Dame, IN 46556, USA}

\author[0000-0002-1172-0754]{Dustin Lang}
\affiliation{Perimeter Institute for Theoretical Physics, Waterloo, ON N2L 2Y5, Canada}

\author[0000-0002-9137-0773]{Paola Marigo}
\affiliation{Department of Physics and Astronomy G. Galilei, University of Padova, Vicolo dell’Osservatorio 3, I-35122, Padova, Italy}

\author[0000-0002-3188-2718]{Hannah Richstein}
\affiliation{Department of Astronomy, University of Virginia, 530 McCormick Road, Charlottesville, VA 22904, USA}

\author[0000-0002-3569-7421]{Edward F. Schlafly}
\affiliation{Space Telescope Science Institute, 3700 San Martin Dr., Baltimore, MD 21218, USA}

\author[0000-0002-2617-5517]{Judy Schmidt}
\affiliation{2817 Rudge Pl, Modesto, CA 95355, USA}

\author[0000-0002-9599-310X]{Erik J. Tollerud}
\affiliation{Space Telescope Science Institute, 3700 San Martin Drive, Baltimore, MD 21218, USA}

\author[0000-0003-1634-4644]{Jack T. Warfield}
\affiliation{Department of Astronomy, University of Virginia, 530 McCormick Road, Charlottesville, VA 22904, USA}

\author[0000-0003-0603-8942]{Andrew Wetzel}
\affiliation{Department of Physics and Astronomy, University of California, Davis, CA 95616, USA}








\begin{abstract}

We present the \jwst\ Resolved Stellar Populations Early Release Science (ERS) science program.  We obtained 27.5~hours of NIRCam and NIRISS imaging of three targets in the Local Group (Milky Way globular cluster M92, ultra-faint dwarf galaxy Draco~{\sc II}, star-forming dwarf galaxy WLM), which span factors of $\sim10^5$ in luminosity, $\sim10^4$ in distance, and $\sim10^5$ in surface brightness. We describe the survey strategy, scientific and technical goals, implementation details, present select NIRCam color-magnitude diagrams (CMDs), and validate the NIRCam exposure time calculator (ETC).  Our CMDs are among the deepest in existence for each class of target. They touch the theoretical hydrogen burning limit in M92 ($<0.08$\,$M_{\odot}$; SNR $\sim5$ at $m_{F090W}\sim28.2$; $M_{F090W}\sim+13.6$), include the lowest-mass stars observed outside the Milky Way in Draco~II (0.09\,$M_{\odot}$; SNR $=10$ at $m_{F090W}\sim29$; $M_{F090W}\sim+12.1$), and reach $\sim1.5$ magnitudes below the oldest main sequence turnoff in WLM (SNR $=10$ at $m_{F090W}\sim29.5$; $M_{F090W}\sim+4.6$). The PARSEC stellar models provide a good qualitative match to the NIRCam CMDs, though are $\sim0.05$~mag too blue compared to M92 F090W$-$F150W data. The NIRCam ETC (v2.0) matches the SNRs based on photon noise from DOLPHOT stellar photometry in uncrowded fields, but the ETC may not be accurate in more crowded fields, similar to what is known for \hst. We release beta versions of DOLPHOT NIRCam and NIRISS modules to the community.  Results from this ERS program will establish \jwst\ as the premier instrument for resolved stellar populations studies for decades to come.
\end{abstract}



\section{Introduction} 
\label{sec:intro}

The resolved stellar populations of nearby galaxies are central to a wide range of astrophysics.  The observed colors, luminosities, and spectral features of resolved stars in galaxies within the Local Volume (LV) anchor our knowledge of star formation \citep[e.g., star cluster formation, the initial mass function, the importance of binarity;][]{massey2003, mckee2007, sarajedini2007, bastian2010, sana2012, kroupa2013, krumholz2014, piotto2015, krumholz2019}, stellar feedback \citep[e.g., the interplay between stars and their immediate surroundings; e.g.,][]{oey1996, dohmpalmer1998, stinson2006, stinson2007, governato2010,  ostriker2010, lopez2011, pellegrini2011, lopez2014, elbadry2016, mcquinn2018, mcquinn2019b}, dust production and characteristics \citep[e.g.,][]{gordon2003, boyer2006, boyer2010, meixner2010, dalcanton2015, schlafly2016, green2019, gordon2021, MericaJones2021}, and stellar chemistry and kinematics across a wide range of environments \citep[e.g.,][]{venn2004, simon2007, geha2009, kirby2011, collins2013, vargas2013,gilbert2014, vargas2014, ji2016a, gilbert2019b, escala2020, kirby2020, gilbert2022}.  Resolved stars are the basis for the local distance ladder \citep[e.g.,][]{freedman2001, Riess:2011fj, beaton2016, riess2016, mcquinn2017, mcquinn2019, freedman2020, riess2021}, which provides constraints on the expansion of the Universe and the nature of dark energy \citep[e.g.,][]{divalentino2021}.  They anchor our knowledge of the stellar evolution models that are used to interpret the light of distant galaxies \citep[e.g.,][]{dotter2008, ekstrom2012, girardi2010, bressan2012, vandenberg2012, choi2016, eldridge2017, conroy2018, hidalgo2018, eldridge2022} and provide detailed insight into the dynamic assembly of our Galactic neighborhood, cosmic reionization, the first stars, near-field cosmology, and the nature of dark matter on the smallest scales \citep[e.g.,][]{mateo1998, tolstoy2009, brown2014, weisz2014a, boylankolchin2015, gallart2015, mcquinn2015, frebel2015, wetzel2016, bullock2017, starkenburg2017a, mcconnachie2018, kallivayalil2018, conroy2019, simon2019, patel2020, boylankolchin2021, sacchi2021, pearson2022}.

Over the past $\sim30$ years, much of this science has been enabled by the \textit{Hubble Space Telescope} (\hst). Since its first images of resolved stars in the local Universe \citep[e.g.,][]{paresce1991, campbell1992, guhathakurta1992, freedman1994, hunter1995}, \hst's exquisite sensitivity, angular resolution, and broad wavelength coverage have transformed our knowledge of the Universe by observing hundreds of nearby galaxies for thousands of hours \citep[e.g.,][]{freedman2001, brown2006, holtzman2006, dalcanton2009, brown2012, dalcanton2012b, gallart2015, riess2016, skillman2017, tully2019, williams2021}, including the Panchromatic Hubble Andromeda Treasury (PHAT) program, which resolved $100$~million stars across the disk of M31 \citep{dalcanton2012, williams2014}.  

However, while \hst\ continues to catalyze new astrophysical insights in nearby galaxies, it has only scratched the surface of science enabled by infrared (IR) observations.  Compared to the UV and optical, \hst's IR camera has coarser angular resolution, which limits it to brighter stars due to stellar crowding, and it can only observe a small portion of the IR spectrum, which limits the types of stellar populations it can study.

\jwst\ will be transformative for resolved stellar populations in the IR. Compared to any other facility, \jwst\ will resolve individual stars at larger distances, to fainter luminosities, over wider color baselines, in more crowded areas, and in regions of higher extinction. \jwst\ can provide the first main-sequence turnoff-based (MSTO) star formation histories (SFHs) of galaxies beyond the Local Group \citep[LG; e.g.,][]{weisz2019c}, systematically measure the sub-Solar mass stellar IMF directly from star counts as a function of environment \citep[e.g.,][]{geha2013, kalirai2013, elbadry2017, gennaro2018a, filion2020, gennaro2020}, determine proper motions and orbital histories for dozens of galaxies outside our immediate Galactic neighborhood \citep[e.g.,][]{vandermarel2012a, sohn2013, kallivayalil2013, zivick2018, sohn2020, warfield2022}, construct parsec-scale maps of the interstellar medium (ISM) in galaxies out to several Mpc \citep[e.g.,][]{dalcanton2015, gordon2016, MericaJones2017, MericaJones2021}, establish a new anchor to the physics of the evolved stars that dominate the rest-frame near-IR light of distant galaxies \citep[e.g.,][]{maraston2006, melbourne2012, boyer2015, boyer2019}, provide high fidelity distances to galaxies throughout the Local Volume \citep[e.g.,][]{beaton2016,mcquinn2019, tully2019,freedman2020,riess2021}, and much more.

With these remarkable capabilities in mind, we have undertaken the \jwst\ Resolved Stellar Populations Early Release Science (ERS) Program (DD-1334; PI D.\ Weisz) to establish \jwst\ as the premier facility for the study of resolved stellar populations in the IR such that it can match and exceed \hst's successes in the local Universe.  To realize this goal, our ERS program has acquired deep multi-band NIRCam and NIRISS imaging of three targets in the Local Group (LG): one Milky Way globular cluster (GC; M92), one ultra-faint dwarf galaxy (UFD; Draco~{\sc II}), and one distant star-forming dwarf galaxy (WLM).  These diverse targets showcase a broad range of the science described above and enable the development and testing of \jwst-specific modules for the widely used crowded field stellar photometry package DOLPHOT \citep{dolphin2000b, dolphin2016}. 

In this paper, we summarize the design of our ERS program, illustrate the new \jwst-specific capabilities of DOLPHOT, outline the photometric reduction process, present a first look at \jwst\ observations of our targets, and undertake select comparisons with stellar models and the current \jwst\ exposure time calculator (ETC). Papers in preparation by our team will provide a detailed overview of the new NIRCam and NIRISS modules for DOLPHOT and will focus on a wide variety of science results enabled by the ERS data beyond what is described here.

This paper is organized as follows.  We summarize the program's overarching science and technical aims and target selection in \S \ref{sec:survey}.  We then describe how we translated these goals into an observational strategy in \S \ref{sec:strategy}.   We detail the actual ERS observations in \S \ref{sec:obs} and summarize the application of DOLPHOT in \S \ref{sec:phot}.  In \S \ref{sec:discuss} we present NIRCam color-magnitude diagrams (CMDs)  and compare them to select stellar models and evaluate the performance of NIRCam ETC.

\section{Program Goals}
\label{sec:survey}

Our team developed a set of main science and technical goals based on anticipated common community use cases of \jwst\ for resolved stellar populations.  For simplicity, we limited our considerations to science cases based on imaging with NIRCam, which is considered the ``workhorse'' camera of \jwst, as well as NIRISS imaging, which we used in parallel.  This setup is analogous to the commonly used mode of \hst\ in which ACS/WFC operates as the primary instrument with WFC3/UVIS acquiring imaging in parallel \citep[e.g.,][]{dalcanton2012b, gallart2015, skillman2017, albers2019, williams2021}.   

Science based on imaging of resolved stars often requires stellar photometry in crowded fields.  Because of that, resolved stellar population studies are technically daunting, requiring highly optimized observations and sophisticated analysis tools that have been developed and refined over the past $\sim40$ years \citep[e.g.,][]{buonanno1979, tody1980, stetson1987, schechter1993, stetson1994, anderson2000, dalcanton2012, williams2014}.  A main technical goal of our program is to develop and release NIRCam and NIRISS modules for DOLPHOT, along with practical recommendations and demonstrations for applying DOLPHOT to NIRCam and NIRISS imaging.  Here, we summarize our main science goals, technical goals, and science ``deliverables'' which guide our ERS program.

\subsection{Scientific Goals}
\label{sec:science}

Our team identified six main science themes that guided the construction of our ERS program.  They are:

\begin{enumerate}

    \item \textit{Star Formation Histories.}  A galaxy's resolved stellar content encodes its star formation history (SFH), which can be reconstructed by fitting CMDs with stellar evolution models \citep[e.g.,][]{tosi1989, tolstoy1996, harris2001, dolphin2002, hidalgo2009}.  These SFHs are particularly robust when CMDs extend below the oldest main sequence turnoff \citep[MSTO; e.g.,][]{gallart2005}.  The faintness of this feature in the optical ($M_V \sim +4$) has limited current `gold standard' SFHs to galaxies within the LG.  
    However, the relatively low effective temperatures of these stars, combined with the decreased sky background in the near-IR and 
     \jwst's excellent sensitivity and angular resolution, will enable it to measure the first SFHs based on the oldest MSTOs for galaxies outside the LG (e.g., \citealt{weisz2019c}; \jwst-GO-1617 PI K.\ McQuinn) from which outstanding questions (e.g., the effects of reionization and/or environment on galaxy formation) can be uniquely addressed \citep[e.g.,][]{bullock2017, simon2019}.  Our \jwst\ program will showcase \jwst's ability to measure robust SFHs.

    \item \textit{The Sub-Solar Mass IMF.}      Resolved star counts shows that lowest-mass galaxies appear to have sub-Solar IMF slopes which deviate from the Galactic value. \citep[e.g.,][]{geha2013, kalirai2013, gennaro2018}.  However, even with \hst, it has proven challenging to acquire sufficiently deep data \citep[down to $\sim 0.2\,\mathmsun$][]{elbadry2017, gennaro2018, gennaro2018a} to unambiguously confirm these putative IMF variations. Our ERS program will illustrate \jwst's capabilities for definitively measuring the sub-Solar IMF in a ultra-faint MW satellite, paving the way for a systematic study of the low-mass IMF and star formation in extreme environments.
    
    \item \textit{Proper Motions.} High-precision astrometry enables the measurement of proper motions (PMs) throughout the LG. \textit{Gaia} has been transformative for objects in the MW halo, while \hst\ has laid the foundation for fainter, more distant systems. \jwst\ is the future of precision astrometry for faint and/or more distant objects. On its own, and in tandem with \textit{Gaia}, \hst, and \textit{Roman}, \jwst\ imaging will provide measurements of total masses, dark matter profiles, and orbital histories for $\sim100$~galaxies in and around the LG \citep[e.g.,][]{bullock2017, kallivayalil2015, fritz2018, gilbert2019,  battaglia2022, warfield2022}.   Our ERS program will showcase the PM measurements capabilities of \jwst. 
    
    \item \textit{Evolved Stars.} Cool evolved stars such as red supergiants and asympotic giant branch (AGB) stars are responsible for 20--70\% of the rest-frame near-IR luminosity of star-forming galaxies  \citep[e.g.,][]{maraston2006, melbourne2012} and are sites of dust production \citep[e.g.,][]{ventura2001}.
    However, the rapid evolution of dusty AGB stars is challenging to model \citep[e.g.,][]{maraston2006, girardi2010, conroy2013, marigo2017}, which has only begun to be alleviated by recent observations \citep[e.g.,][]{boyer2015, boyer2017}.  \jwst's expansive IR filter set will reveal elusive dust-enshrouded populations of AGB stars (e.g., oxygen-rich M stars and carbon-rich C stars) across a wide range of galactic environments \citep[e.g.,][]{hjort2016, jones2017, marini2020}. Our ERS program will demonstrate \jwst's capacity to study IR-bright, evolved stars.

    \item \textit{Extinction Mapping.} In the LG, \textit{Spitzer} and \textit{Herschel} have mapped dust emission at $\sim10-40$\arcsec\ and $\sim7-12$\arcsec\ resolution, respectively \citep[10~pc for the Magellanic Clouds; 100~pc for M31 and M33;][]{draine2007, gordon2014, chastenet2019, utomo2019}.  \jwst\ can map the cold ISM at significantly higher spatial resolution by inferring dust content from its impact on stellar spectral energy distributions \citep[e.g.,][]{dalcanton2015, gordon2016}.  Our ERS observations will demonstrate \jwst's ability to map dust extinction and relate it to properties of the cold ISM.

    \item \textit{Ages of Globular Clusters.} Accurate ages of the oldest GCs are particularly important for connecting the stellar fossil record to events in the early Universe including cosmic reionization and the age of the Universe itself \citep[e.g.,][]{chaboyer1996, grebel2004, ricotti2005, monelli2010b, brown2014, weisz2014b, boylankolchin2015}.  Current age estimates are typically limited to $\sim1$~Gyr precision (i.e., twice as long as reionization lasted) due to the age-metallicity degeneracies at the MSTO (see \citealt{boylankolchin2021} and references therein). \jwst\ observations of the `kink' on the lower main sequence  (MS) can yield more precise estimates of cluster ages \citep[e.g.,][]{sarajedini2009, bono2010, kalirai2012, correnti2018}. Our ERS data will showcase the powerful capabilities of \jwst\ for precise GC age-dating.

\end{enumerate}

Beyond enabling our main science goals, we sought to identify observations that would make our ERS program rich for archival pursuits.  Examples include measuring extragalactic distances in \jwst\ bands \citep[e.g., TRGB, variable stars;][]{beaton2016, madore2018, mcquinn2019}, identifying rare stars from low-mass metal-poor stars to luminous red supergiants \citep[e.g.,][]{schlaufman2014, casey2015, levesque2018}, searching for dust production among red giant branch stars \citep[RGB; e.g.,][]{boyer2006, boyer2010}, and examining the nature of dark matter using wide binaries \citep[e.g.,][]{penarrubia2016}.

\subsection{Technical Goals}
\label{sec:technical_goals}

The main technical goal of our ERS program is to enable resolved star science by the broader community.  At the heart of this goal is the addition of NIRCam and NIRISS modules to DOLPHOT.  This process includes the technical development of NIRCam and NIRISS modules for DOLPHOT, testing their performance on real data, releasing data products that immediately enable science (e.g., stellar catalogs), and providing guidance to the community on best use practices of DOLPHOT for future applications.  Here, we broadly describe each of these technical goals and how they influenced the observational strategy of our ERS program.

As with previous updates to DOLPHOT \citep[e.g.,][and many unpublished updates]{dalcanton2012b, dolphin2016}, the core functionality of the code remains the same as described in \cite{dolphin2000b}, but certain aspects have been updated for NIRCam and NIRISS.  

The DOLPHOT modules for NIRCam and NIRISS each feature their own pre-processing routines that apply masks (e.g., of reference, saturated, and other unusable pixels) to the images based on the data quality flags provided by the STScI pipeline.  They also apply the pixel area maps appropriate to each camera.  Other updates include the use of photometric calibrations provided in the image metadata, conversions to VEGAMAG, and camera-specific PSF models with corresponding encircled energy corrections.

For testing DOLPHOT on real data, we identified several observational scenarios that we anticipate to be common for NIRCam and NIRISS studies of resolved stars. They are:

\begin{enumerate}
    \item Targets with various levels of crowding.  This includes images that are completely uncrowded (e.g., in which aperture vs PSF photometry can be compared), images with highly variable amounts of crowding (e.g., due to surface brightness variations), and highly crowded images (i.e., the photometric depth is primarily limited by crowding).  
    \item Targets that include stars spanning a large dynamic range in brightness in the same image.  An example would be a GC, in which there are very bright red giants and extremely faint dwarfs.  This enables a variety of tests, including the ability to recover faint sources next to very bright objects.
    \item Targets with bright, saturated stars.  \jwst\ is extremely sensitive.  Understanding the degree to which saturated stars affect the photometry of fainter objects will be important to a variety of science goals.
    \item Targets that demonstrate the ability of using the higher angular resolution short-wavelength (SW) images to increase the accuracy of the long-wavelength (LW) photometry.  PHAT showed that joint reduction of \hst\ optical and IR data produced IR photometry that provides significantly sharper CMDs compared to reducing IR data alone \citep[e.g.,][]{williams2014}.  Similar gains should be possible with NIRCam.

    \item Targets that enable the simultaneous reduction of \hst\ and \jwst\ imaging.  To date, DOLPHOT has produced wonderful cross-camera results for \hst\ \citep[e.g.,][]{dalcanton2012, williams2014, williams2021}, but it needs to be vetted and optimized for cross-facility use.

\end{enumerate}

\begin{figure*}
\epsscale{1.18}
\plotone{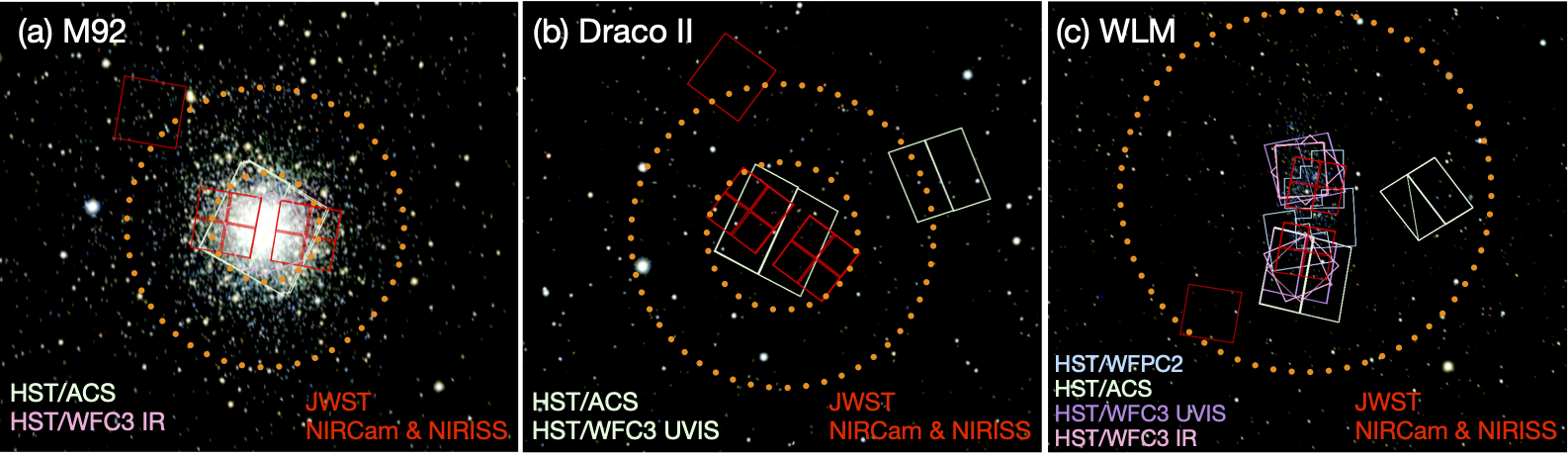}
\caption{The locations of our NIRcam and NIRISS observations (plotted in red) for each ERS target, overplotted on a Pan-STARRS optical image.   The orange dotted lines indicate: (a) 2 and 5 half-light radii ($r_{\rm h}$),  (b) 1 and 2\,$r_{\rm h}$ and (c) 1\,$r_{\rm h}$ of  each target.    We show additional pointings for each system:   (a) M92:   select HST optical (green; HST-GO-10775; Sarajedini et al. 2007) and IR (pink; HST-GO-11664; e.g., Brown et al. 2010).    (b) \dracoii:  HST/ACS optical data (HST-GO-14734; PI Kallivayalil) are shown in green.  (c) WLM:   An exhaustive, though not complete set of HST observations including HST/WFPC2 UV and optical imaging in blue (HST-GO-11079; Bianchi et al. 2012), HST/WFC3 UVIS UV imaging in purple (HST-GO-15275 PI Gilbert), HST/ACS and HST/UVIS optical imaging in green (HST-GO-13768; PI Weisz, Albers et al. 2019), and HST/WFC3 IR imaging in pink (HST-GO-16162, PI Boyer).   We opted not to undertake large dithers to fill the NIRCam chip and module gaps which would have substantially increased the program time while only marginally enhancing our science goals. \label{fig:footprints}}
\end{figure*}

\subsection{Deliverables}

Our program is in the process of providing several ``deliverables'' to the community that can be found on our team website\footnote{\url{https://ers-stars.github.io}}.  A primary deliverable is the public release of DOLPHOT with NIRCam and NIRISS specific modules for which ``beta'' versions can be found on the main DOLPHOT website\footnote{\url{http://americano.dolphinsim.com/dolphot/}}. This software enables crowded field stellar photometry for a diverse range of science in the local Universe.  Along with the software release we will provide extensive documentation of how to use DOLPHOT and examples of it applied to our ERS observations.  Following careful calibration and testing, we will release high level science products including the output of our team DOLPHOT runs on ERS data (e.g., diagnostic plots and files), and NIRCam and NIRISS stellar catalogs for each target along with artificial star tests.  These data products will be refined as our understanding of \jwst\ improves (e.g., due to updated PSF models) and will eventually include examples of how to use DOLPHOT for simultaneous reduction of \hst\ and \jwst\ imaging.

\begin{figure*}
\plotone{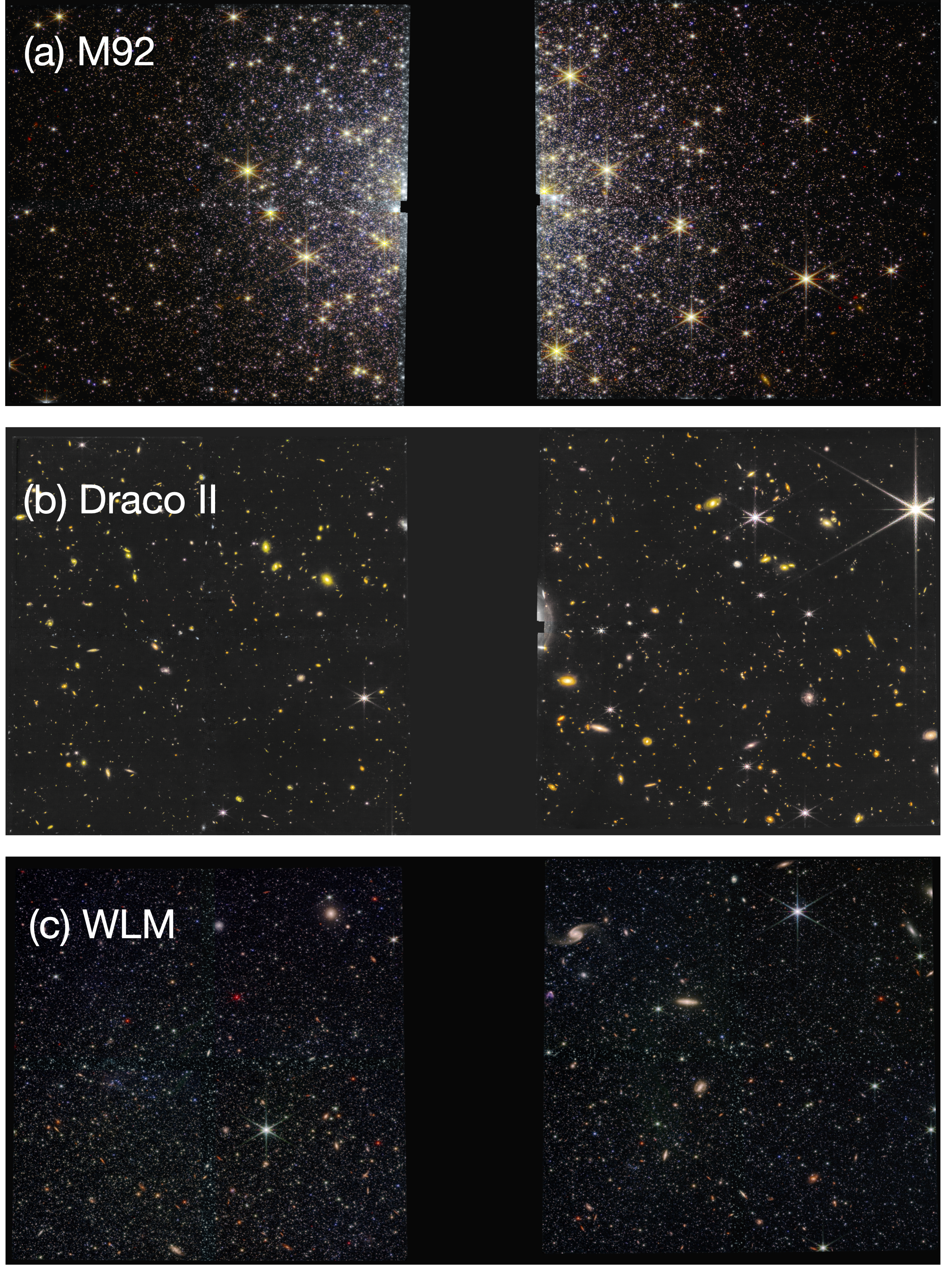}
\caption{NIRCam color composite images for our 3 ERS targets.   In each RGB image, F090W was used as the blue channel, F150W as the green and a combination of the two LW filters as the red channel. \label{fig:m92_color}}
\end{figure*}

\section{ Strategy}
\label{sec:strategy}

\subsection{Filters}

The diversity of our science cases required careful consideration of filter selection.  Several of our science goals are centered around maximizing depth, color baseline, and astrometric precision.  Accordingly, we primarily focused on SW filter selection, which has better sensitivity (for most stars) and angular resolving power than the long-wavelength channel.   

Using an ancient, metal-poor isochrone (12.5~Gyr, [Fe/H]=$-2.0$) from the MIST stellar models \citep{choi2016}, we examined the expected performance for the SW wide filter (F070W, F090W, F115W, F150W, F200W) permutations at three different CMD locations: the blue HB ($\Teff\sim7000$~K), the MSTO ($\Teff\sim6000$~K), and the lower MS ($\sim0.2\,\mathmsun$; $\Teff\sim4000$~K).   At each point, we used the pre-commissioning \jwst\ ETC (v1.1.1) to compute the exposure time required to reach a SNR$=10$ for the ``scene'' in the ETC.

The best performing filters for our areas of consideration are F090W, F115W, and F150W.  They all exhibit comparable performance at the HB and MSTO.  However, F090W requires 2.5 times more exposure time to achieve the same SNR for a $0.2\,\mathmsun$ star as either F115W or F150W.  Nevertheless, we opted for F090W over F115W because compared to F115W$-$F150W, F090W$-$F150W provides superior color information for most stars and F090W has the potential for higher angular resolution (if dithered appropriately), which is critical for astrometry.  Finally, the similarity between F090W and \hst/F814W (or Johnson I-band) provides useful features such as matching catalogs between facilities and TRGB distance determinations \citep[e.g.,][]{mcquinn2019}. F070W and F200W provide the largest color baseline, but each filter is less sensitive to stars far from their effective wavelength.  For example, F070W required 4 times more integration time for a $0.2\,\mathmsun$ star than the next bluest filter, F090W.  F200W requires twice as much exposure time for a HB star than F150W.

We opted to use the same F090W$-$F150W filter combination for all targets to provide for an empirical comparison between the GC and UFD \citep[e.g.,][]{brown2012} and for good sampling of the oldest MSTO in the distant dwarf.   We considered more than two SW filters, but the cost of acquiring extra data outweighed the scientific utility.  We selected simultaneously observed LW filters on a per target basis, as they enable secondary science unique to each object.  Finally, we selected F090W and F150W for parallel NIRISS imaging for consistency with NIRCam.

We emphasize that while our filter combinations represent a good compromise across the CMD for our program goals, they may not be optimal for all science cases. We encourage exploration tailored to a program's particular science aims.

\subsection{Target Selection}

We selected targets by first considering all known GCs in the MW \citep{harris2010} and galaxies within $\sim1$~Mpc \citep{mcconnachie2012}, including updates to both catalogs and discoveries through 2017 \citep[e.g.,][]{laevens2014, laevens2015, bechtol2015, drlicawagner2015, koposov2015}.  The limiting distance was selected to ensure we could reach the oldest MSTO with SNR$=10$ in the most distant system in a reasonable amount of time based on previous experience with \hst\ \citep[e.g.,][]{cole2014, albers2019} and results from the \jwst\ exposure time calculator (ETC).

We required that each target have extensive \hst\ imaging (e.g., to enable combined \hst\ and \jwst\ proper motion studies, create panchromatic stellar catalogs) and have a good sampling of ground-based spectra (e.g., for full phase space information, comparing stellar properties from spectra and photometry, incorporating stellar abundance patterns into various analyses).  

We then identified a minimum set of targets that could be used to achieve our science and technical goals: one MW GC, one UFD, and one more distant star-forming dwarf galaxy.  We then sought to maximize observational efficiency by focusing on some of the nearest examples of these classes.  We eliminated targets that were not visible during the nominal ERS window. 


This selection process yielded three targets: MW GC M92, MW satellite UFD \dracoii, and star-forming dwarf galaxy WLM.  Basic observational characteristics of these targets are listed in Table \ref{tab:obs}.  We detail the observational strategy for each target in the following sections.

\begin{table}
    
    \caption{Basic observational properties of the three ERS targets. Properties for M92 have been taken from the updated MW GC catalog of \citet{harris2010}, while those of \dracoii\ and WLM are from the updated LG galaxy catalog of \citet{mcconnachie2012}. Note that $\mu_0$ is the effective surface brightness and $r_h$ is the half-light radius.  \label{tab:obs}}
    \centering
    \begin{tabular}{lccc}

    \toprule
     & M92 & \dracoii\ & WLM \\
    \toprule
    
    RA (J2000) & 17h17m07.27s & 15h52m47.60s & 00h01m58.16s \\
    Dec (J2000) & $+$43d08m11.5s & $+$64d33m55.0s & $-$15d27m39.34s \\
    $M_V$ (mag) & $-8.2$  & $-0.8$ & $-14.2$ \\
    E(B-V) (mag) & $0.02$ & $0.01$ & $0.03$ \\
    $(m-M)_0$ (mag) & $14.6$ & $16.9$ & $24.9$ \\
    $\mu_0$ (mag arcsec$^{-2}$) & $15.5$ & $28.1$ & $24.8$ \\
    $r_h$ (\arcmin) & $1.0$ & $2.7$ & $7.8$ \\
    
     \hline
    \toprule

    \end{tabular}
\end{table}

\subsection{M92}
\label{sec:m92}

M92 (NGC~6341) is a well-studied, metal-poor GC in the MW that is often used as a benchmark for extragalactic stellar population studies and for photometric calibration (e.g., to verify zero points; e.g., \citealt{dalcanton2009, brown2014, gallart2015}). Imaging this system satisfies several science and technical goals including GC ages, individual star proper motions, the present day mass function, testing DOLPHOT over a large dynamic range of stellar brightness and spatially varying stellar density, and gauging the effects of bright saturated stars on the photometric process.  

As illustrated in Figure~\ref{fig:footprints}, we placed the NIRCam field near the center of M92, with the aim of maximizing NIRCam spatial overlap with a wealth of multi-band \hst\ imaging of M92.  The parallel NIRISS field is located at $\sim5$ half-light radii.  We constrained the orientation such that the NIRISS field had a modest probability of overlapping at least some \hst\ data in the outer regions.  However, orientations allowed by the final ERS window did not result in overlap between NIRISS and \hst\ imaging.   

We chose the F090W, F150W, F277W, and F444W for our NIRCam imaging of M92.  We selected F277W and F444W for their broad scientific utility including studying the lower MS kink at long wavelengths \citep[e.g.,][]{sarajedini2009, bono2010}, searching for dust production at low-metallicities \citep[e.g.,][]{boyer2006, boyer2010}, and exploring multiple populations in the IR \citep[e.g.,][]{milone2012, milone2014, correnti2016, milone2017}.

We aimed to reach SNR$=10$ at $0.1\,\mathmsun$ in F090W and F150W, which we estimate to be $m_{F090W} \approx26$ ($M_{F090W} \approx+11.4$) and $m_{F150W} \approx25.8$ ($M_{F150W} \approx+11.2$) based on the MIST isochrones and the characteristics of M92 listed in Table~\ref{tab:obs}. The 2017 versions of the ETC (v1.1.1) and APT (v25.1.1) yielded exposure times of 1288s in each filter.  The anticipated SNRs in the LW filters at $m_{F090W} \approx26$ were $21.1$ in F277W and 6.4 in F444W. We estimated NIRISS imaging to be marginally shallower than NIRCam with integration times of 1074s in each of F090W and F150W.  The difference in duration is set by facility overheads as calculated by APT.

\subsection{Draco {\sc II}}

At $\sim$20~kpc, \dracoii\ is one of the nearest examples of a MW satellite \citep{laevens2015b, longeard2018}.  At the time of program design, \dracoii\ was designated as a UFD \citep[i.e., it has dark matter;][]{willman2012, martin2016}.  In the interim, there has been some debate in the literature over its status as a UFD vs GC \citep[e.g., whether or not it has dark matter, a metallicity spread, the slope of its sub-Solar stellar mass function; e.g.,][]{longeard2018, baumgardt2022}, an issue our \jwst\ data should help resolve.  

\dracoii's close proximity provides for efficient \jwst\ imaging that reaches far down the lower main sequence ($\lesssim0.2\,\mathmsun$).  Our deep \dracoii\ data satisfies several of our science goals including measuring the low-mass stellar IMF (which can shed insight into its status as a GC or UFD), determining the SFH of an ancient sparsely populated system, measuring proper motions of individual faint stars using \hst\ and \jwst, and exploring our ability to distinguish between faint stars and unresolved background galaxies.

We placed the NIRCam field on the center of \dracoii.  The field overlaps with archival \hst\ imaging obtained in March 2017 (GO-14734; PI N.\ Kallivayalil), Keck spectroscopy, and is well-matched to the half-light radius.  To maximize scheduling opportunities, we did not constrain the orientation.  The NIRISS field is unlikely to contain many \dracoii\ member stars, so its exact placement was not crucial.  The primary use of the NIRISS data will be to aid with modeling contamination (e.g., foreground stars and background galaxies) in the F090W$-$F150W CMD, e.g., as part of measuring the low-mass IMF.

We selected the F090W, F150W, F360M, and F480M for our NIRCam imaging of \dracoii.  The rationale for the two SW filters is described in \S \ref{sec:m92}.  The LW filters are located near metallicity sensitive molecular features in the mid-IR, and they may be suitable for measuring photometric metallicities of metal-poor stars \citep[e.g.,][]{schlaufman2014, casey2015} similar to what is possible in the optical using, for example, the Calcium H\&K lines \citep[e.g.,][]{starkenburg2017a, fu2022}.  For NIRISS, we used the F090W and F150W filters.

Our target depth is set by low-mass IMF science.  Tightly constraining the low-mass IMF in \dracoii\ requires reaching stars $\lesssim0.2\,\mathmsun$ \citep[e.g.,][]{elbadry2017} with SNR$=10$ in F090W and F150W.  Using the MIST stellar models and the observational properites of \dracoii\ listed in Table 1, our target depths are  SNR$\sim10$ at $m_{F090W} =27$ ($M_{F090W} =+10.3$) and $m_{F150W} =26.8$ ($M_{F150W} =+10.1$). Using the 2017 versions of the ETC (v1.1.1) and APT (v25.1.1) we found exposure times of 12798s in F090W and 6399s in F150W would each these depths. We opted on SW/LW combinations of F090W/F480M and F150W/F360M, which provided for SNRs of $\sim7$ (F360M) and $\sim4$ (F480M) at F150W=25.  We estimated the NIRISS imaging to have 12540s in F090W and 6270s in F150W, which will result in marginally shallower CMDs than NIRCam.

\subsection{WLM}
\label{sec:wlm}

WLM \citep{wolf1909, melotte1926} is a metal-poor ([Fe/H]$= -1.2$; \citealt{leaman2009}) star-forming dwarf galaxy at $\sim0.9$~Mpc \citep[e.g.,][]{albers2019}.  Though slightly closer objects of this class exist (e.g., IC~1613), WLM is the nearest example of a low-metallicity environment in which resolved CO clouds have been detected \citep[][]{rubio2015}.  It has a sufficiently high star formation rate in the past several Gyr that it should host a sizable population of AGB stars \citep[e.g.,][]{dolphin2000b, weisz2014a, mcquinn2017b, albers2019}.  In terms of science, our \jwst\ imaging of WLM will allow us to measure its SFH from the ancient MSTO and compare it to the existing \hst-based SFH, measure its bulk PM using archival \hst\ imaging, explore the stellar populations associated with it CO clouds, construct parsec-scale extinction using IR-only techniques and combined UV-optical-IR \hst\ and \jwst\ stellar SEDs \citep[e.g.,][]{dalcanton2015, gordon2016}.  On the technical side, our observations of WLM allow us to test DOLPHOT in a regime of faint, crowded stars, that is typical of more distant systems, and also test its capabilities for simultaneously measuring stellar photometry across facilities (\hst\ and \jwst) and instruments (WFPC2, ACS, UVIS, NIRCam).

We required that one module of the NIRCam observations overlap UV-optical-IR \hst\ observations as well as the ALMA-detected CO clouds \citep{rubio2015}.  The other NIRCam module overlaps with the deep optical \hst/ACS imaging presented in \citet{albers2019}.  We enforced this configuration by requiring orientations of 70--110$^{\circ}$ or 250--290$^{\circ}$.  This orientation placed the NIRISS field in the stellar halo of WLM, providing an expansion on the areal coverage to build on the population gradient studies in WLM \citep[e.g.,][]{leaman2013, albers2019}.  We used artificial star tests associated with the deep \hst\ imaging of \citet{albers2019} to ensure that our NIRCam observations would only be modestly affected by stellar crowding. 

We selected the F090W, F150W, F250M, and F430M for our NIRCam imaging of WLM.  As discussed in \S \ref{sec:m92}, the SW filter combination F090W$-$F150W is likely to be widely used for SFHs measured from the ancient MSTO.  Medium band filters in the near-IR have proven remarkably efficient for photometric identification and classification of AGB stars \citep[e.g.,][]{boyer2017}.  Our simulations, based on these studies, suggest that the F250M and F430M filters should work well for similar science.    For NIRISS, we selected the F090W and F150W filters.

In the optical, measuring a well-constrained SFH for distant dwarf galaxies requires a CMD that reaches a SNR$\sim$5--10 at the oldest MSTO  \citep[e.g.,][]{cole2007, monelli2010b, cole2014, skillman2014, gallart2015, skillman2017, albers2019}.  Using the MIST stellar models, we find that the ancient MSTO for metal-poor stellar population has $M_{F090W} =+3.4$ and $M_{F150W} =+3.1$.  Using the parameters for WLM in Table \ref{tab:obs} and v1.1.1 of the ETC, we estimated that 33241s in F090W and 18724s in F150W would reach the required depths of $m_{F090W} =28.3$ and $m_{F150W} =28.1$ with SNR=10 in each filter.

\begin{table*}
    
    \caption{A summary of our \jwst\ ERS observations taken in 2022. \label{tab:jwst_obs}}
    \centering
    \begin{tabular}{lccccccc}

    \toprule
    Target & Date & Camera & Filter & $t_{exp}$ [s] & Groups & Integrations & Dithers\\
    \toprule
    M92 & June 20--21 & NIRCam & F090W/F277W & 1245.465 & 6 & 1 & 4\\
     & & NIRCam & F150W/F444W & 1245.465 & 6 & 1 & 4\\
     & & NIRISS & F090W & 1245.465 & 7 & 1 & 4\\
     & & NIRISS & F150W & 1245.465 & 7 & 1 & 4\\
     Draco~{\sc II} & July 3 & NIRCam & F090W/F480M & 11810.447 & 7 & 4 & 4\\
     & & NIRCam & F150W/F360M & 5883.75 & 7 & 2 & 4\\
     & & NIRISS & F090W & 11123.294 & 9 & 7 & 4\\
     & & NIRISS & F150W & 5883.75 & 10 & 3 & 4\\
     WLM & July 23--24 & NIRCam & F090W/F430M & 30492.427 & 8 & 9 & 4\\
     & & NIRCam & F150W/F250M & 23706.788 & 8 & 7 & 4\\
     & & NIRISS & F090W & 26670.137 & 17 & 9 & 4\\
     & & NIRISS & F150W & 19841.551 & 19 & 6 & 4\\
     \hline
    \toprule

    \end{tabular}
\end{table*}

\subsection{Program Updates Since 2017}

Our program has only had minor changes since it was first approved in November 2017.  In 2021, we changed the DEEP2 readout patterns for WLM and \dracoii\ to MEDIUM8, following updated advice from a STScI technical review.  The number of groups and integrations was changed accordingly.  In 2021, STScI increased our allocated time from 27.35~hr to 27.5~hr to reflect updated overhead accounting. 

Following commissioning in spring 2022, STScI staff changed the range of aperture PA ranges of \dracoii\ from unconstrained to be have a value of 47--118$^\circ$, 143--280$^\circ$, or 354--24$^\circ$ in order to avoid the ``claws'' feature\footnote{url{https://jwst-docs.stsci.edu/jwst-near-infrared-camera/nircam-features-and-caveats/nircam-claws-and-wisps}} which is due to stray light from the bright source that is outside the field of view of NIRCam \citep{jdox2016, rigby2022}.

\section{Observations}
\label{sec:obs}

We acquired NIRCam and NIRISS imaging of our three targets in July and August of 2022.  Figure~\ref{fig:footprints} shows the NIRCam and NIRISS footprints for each of our targets overlaid on ground-based images and Table \ref{tab:obs} lists our observational configurations.  Full details on implementation can be viewed by retrieving proposal 1334 in the Astronomer's Proposal Tool (APT)\footnote{\url{apt.stsci.edu}}.   

We followed the same fundamental observing strategy for all three targets.  For each target, the primary instrument, NIRCam, imaged a single central field in SW and LW filters as described in \S \ref{sec:obs}. NIRISS obtained imaging in parallel in the two filters F090W and F150W.

All observations were taken with the 4 point subpixel dither pattern \texttt{4-POINT-MEDIUM-WITH-NIRISS}.  This pattern ensured adequate PSF sampling for both cameras, as well as improved rejection of cosmic rays, hot pixels, etc.  

We opted against primary dithers.  The main advantage of the primary dithers is to fill the gaps between the dectectors and/or modules.  However, the inter-gap imaging is generally shallower than the rest of the data and requires more \jwst\ time to acquire.  For our particular science cases, including primary dithers added a modest amount of time to the program but would not substantially enhance the data for our main science goals. Other science cases (e.g., covering a large region such as PHAT did) may benefit from primary dithers and filling gaps.

For M92, we used the \texttt{SHALLOW4} readout pattern for NIRCam and the \texttt{NIS} readout pattern for NIRISS.  The orientation was constrained to an Aperture PA range of 156 to 226$^\circ$ in order to maximize the possibility of overlap between the NIRISS field and existing \hst\ imaging, while also allowing for reasonable schedulability very early in the lifetime of \jwst.  The central location of the NIRCam field ensures it will overlap with existing \hst\ imaging.  Including overheads, the total time charged for observing M92 was $7037s$.  The ratio of science to charged time was $\sim0.35$ (counting the primary NIRCam imaging only) or $\sim0.7$ (for both NIRCam and NIRISS imaging).   The data volume was well within the allowable range.

For Draco~{\sc II}, we used the \texttt{MEDIUM8} readout pattern for NIRCam and the \texttt{NIS} readout pattern for NIRISS.  The NIRCam field was centered on the galaxy and will largely overlap with existing \hst\ imaging and Keck spectroscopic data.  The large angular separation of the NIRISS and NIRCam fields compared to the size of Draco~{\sc II} means that the NIRISS field will contain few, if any, \textit{bona fide} members of Draco~{\sc II}.  Including overheads, the total time charged for observing Draco~{\sc II} was $24539s$.  The ratio of science to charged time was $\sim0.71$ (counting the primary NIRCam imaging only) or $\sim1.36$ (for both NIRCam and NIRISS imaging). The data volume was well within the allowable range.

For WLM, we used the \texttt{MEDIUM8} readout pattern for NIRCam and the \texttt{NIS} readout pattern for NIRISS.   The NIRCam field was placed in the center of WLM in order to overlap the low-metallicity molecular clouds discovered by ALMA as well as deep archival optical HST imaging. Subsequent UV and near-IR \hst\ imaging of WLM obtained by members of our team were placed to maximize the chances of overlap with our \jwst\ observations of WLM.  

Including overheads, the total time charged for observing WLM was $66884s$.  The ratio of science to charged time was $\sim0.8$ (counting the primary NIRCam imaging only) or $\sim1.50$ (for both NIRCam and NIRISS imaging).  The large data volume (19.962~GB) for WLM generated a ``Data Excess over Lower Threshold'' warning in APT.  Though this level of warning is only a recommendation to mitigate data volume excess, we nevertheless consulted with STScI about mitigation strategies.  However, we were unable to identify a way to reduce data volume without compromising the science goals and no changes were made.  We caution that even longer integrations (e.g., that may be needed for deep CMDs outside the LG) may require careful planning to avoid data volume limitations. 

Because our WLM observations span several continuous hours (Table \ref{tab:jwst_obs}),  our observations of WLM should allow us to recover the light curves of short period variables (e.g., RR Lyrae).  Generation of the light curves requires performing photometry on calibrated images at each integration.  At the time of this paper's writing, due to the large data volume, the STScI \jwst\ reduction pipeline generates integration-level calibrated images only when the time series observation (TSO) mode is used.  However, we were unable to use the TSO mode of \jwst\ as it does not permit dithers nor parallel observations.  Instead, producing the necessary images to generate light curves requires running the reduction pipeline locally and creating custom time series analysis software, which is beyond the scope of this paper.  

In total, all of our science observations with NIRCam total 20.45~hr and observations with NIRISS total 17.85~hr, indicating a fairly high ratio of science-to-charged time.

\begin{figure*}[ht!]
\plotone{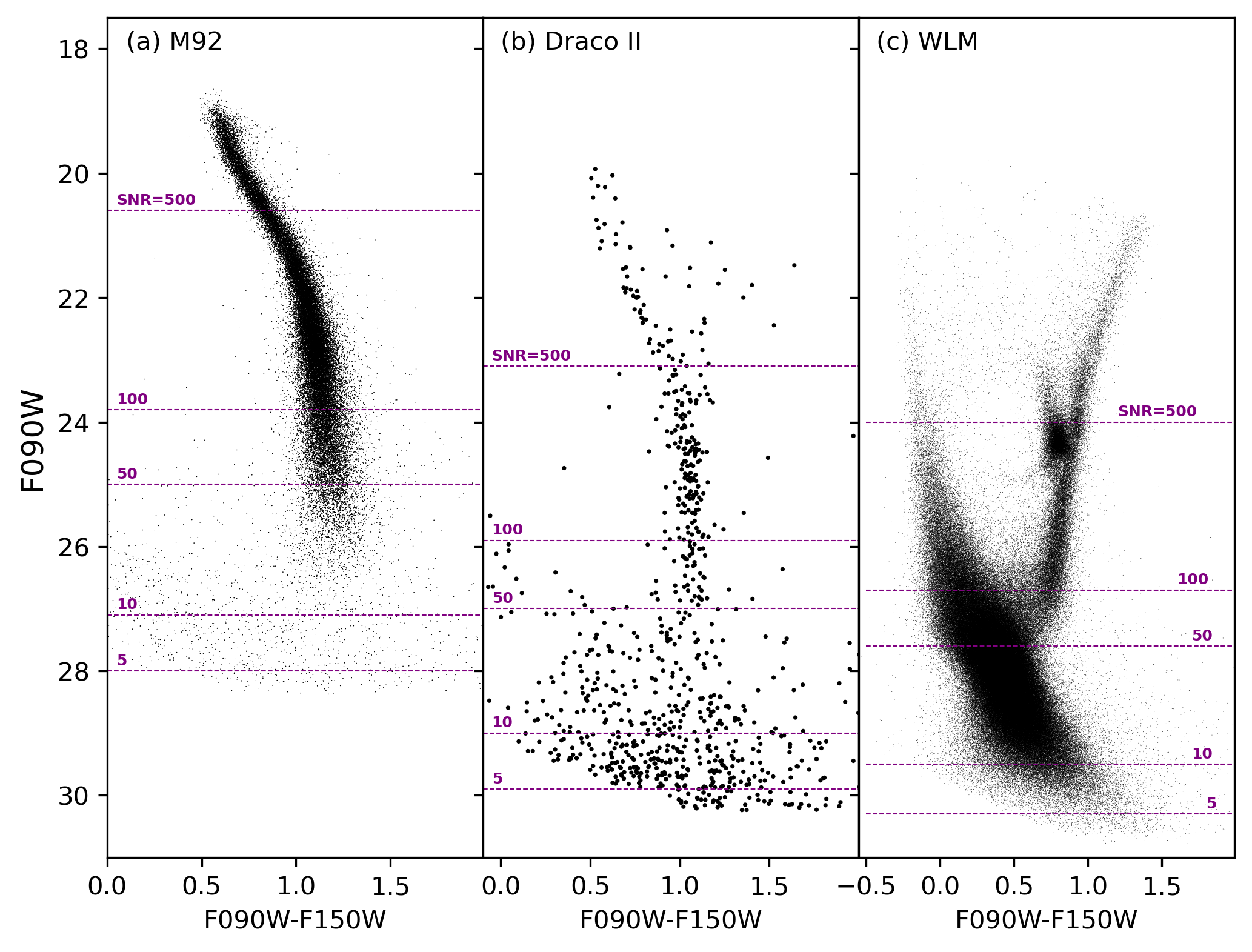}
\caption{CMDs from all 8 NIRCam chips for each of our targets. \textbf{Panel (a):} M92 extends from our bright saturation limit without Frame 0 data (F090W$\sim19$) just below the MSTO and reaches just fainter than the hydrogen burning limit ($F090W>28.2$).  The inflection point of the MS at $F090W\sim21$ is the MS kink. \textbf{Panel (b):} The CMD of \dracoii\ includes the MS kink at $F090W\sim23$ and reaches the bottom of the stellar sequence provided by the standard PARSEC models ($0.09\,\mathmsun$).   \textbf{Panel (c):}  The CMD of WLM displays a wide variety of stellar sequences that span a range of ages including young stars (e.g., the upper main sequence, red core helium burning stars), intermediate age stars (e.g., AGB and RGB stars), and ancient stars that span a wide range of colors and magnitude (e.g., RGB, SGB, oMSTO, lower MS).  This CMD of WLM (SNR$=$10 at $M_{F090W} = +4.6$) is the deepest taken for a galaxy outside the immediate vicinity of the MW. \label{fig:all_SW_CMDs}}
\end{figure*}

\section{Photometry}
\label{sec:phot}
We perform photometric reductions of our observations using the newly developed NIRCam module for DOLPHOT. The detailed description of how DOLPHOT works is well-documented in the literature \citep[e.g.,][]{dolphin2000b,dalcanton2012b, dolphin2016} and the functionality of the new NIRCam module are summarized in \S~\ref{sec:technical_goals}.  A more detailed write up of the DOLPHOT NIRCam and NIRISS modules are the subjects of an upcoming paper from our team.  For this paper, we only focus on the NIRCam data.  The WebbPSF\footnote{\url{https://webbpsf.readthedocs.io/en/latest/}} NIRISS PSF models appear to concentrate light significantly more than what we have observed in the ERS images.  Consequently, our NIRISS photometry is not yet reliable and further updates await improvements to the PSF models.

We first acquired all images from MAST. Per their FITS headers, versions of all \jwst\ images used in this paper have  the following \jwst\ pipeline versioning information \texttt{CAL\_VER}$=$1.7.2,\texttt{CRDS\_VER}$=$11.16.11, and \texttt{CRDS\_CTX}$=$jwst\_p1009.pmap. 
Next, we performed DOLPHOT reductions on the level 2b \texttt{crf} frames and use the level 3 \texttt{i2d} F150W drizzled image as the astrometric reference frame.  The use of this reference image ensures excellent internal alignment of our image stack and ties the absolute astrometry to \textit{Gaia} DR2. We perform photomery in all four bands simultaneously. Since there is no spatial overlap between the footprint of the two NIRCam modules, we photometer them independently and merge the catalogs \textit{a posteriori}. 

We have not included Frame 0 in the photomeric reductions for this paper.  As of this paper's writing, the \jwst\ pipeline does not automatically provide Frame 0, requiring these images be generated locally.  Even when created with our own execution of the \jwst\ pipeline, we have yet to fully resolve how $1/f$ noise issue \citep[e.g.,][]{schlawin2020, bagley2022} can be resolved in a manner that ensures self-consistent photometry with DOLPHOT for all frames.  This issue will be addressed in more detail in our forthcoming photometry paper.  In the meantime, without Frame 0 data, our photometry saturates at fainter magnitudes that we expect to recover once the Frame 0 data in included in our DOLPHOT runs.

For our photometric reductions, we adopt the DOLPHOT parameter setup recommended by PHAT \citep{williams2014}. We use the parameters recommend for ACS from PHAT for the SW images and the WFC3/IR parameters for the LW images. Subsequent data releases will make use of a set of parameters that we are tailoring to NIRCam and NIRISS observations of resolved stars.

We make use of NIRCam PSF models generated with WebbPSF \citep{Perrin2014}. We use the Optical Path Delay maps from July 24th, 2022 (the best matching file to our WLM observation epoch). Inspection of \jwst's wavefront field at the epochs of our three observations show little variation in the optical performance of the telescope, justifying our choice of a common PSF library. We are currently working in quantifying the full effect of \jwst\ time-dependent PSF variations on DOLPHOT photometry.

\begin{table}[]
\caption{Quality-metric criteria used to cull our DOLPHOT photometric catalogs. \label{tab:cuts}}
\centering
\begin{tabular}{lcccccc}
\toprule
Band & SNR & $Sharp^2$ & $Crowd$& $Flag$& $Object \, Type$\\
\hline
F090W & $\geq 4$ & $\leq 0.01$ & $\leq 0.5$ & $\leq 2$ & $\leq 1$\\
F150W & $\geq 4$ & $\leq 0.01$ & $\leq 0.5$ & $\leq 2$ & $\leq 1$\\
\toprule
\end{tabular}
\end{table}

The final products of the DOLPHOT photometric run are catalogs with positions, VEGAmag magnitudes (calibrated using the latests NIRCam zero-points\footnote{\url{https://jwst-docs.stsci.edu/jwst-near-infrared-camera/nircam-performance/nircam-absolute-flux-calibration-and-zeropoints}}), uncertainties based on photon noise, and a set of quality metrics related to the goodness of point-source photometry (e.g., SNR, $\chi^2$ of the PSF fit, angular extent of the source, crowding level). At this stage, metrics such as the photometric error and the SNR are based on the Poissonian treatment of photon noise. While in many cases this is sufficient for rough estimation, there are caveats associated with this approximation, especially when measuring stars in very crowded fields, or close to the limiting magnitude. We will provide a more thorough discussion on SNRs estimation in our upcoming \jwst\ DOLPHOT photometry paper.

The catalogs provided by DOLPHOT are subsequently inspected and culled to remove contaminants (e.g., artifacts, cosmic rays, extended sources) while aiming to retain the largest number of \textit{bona fide} stars. We identify a set of quality-metric cuts, listed in Table~\ref{tab:cuts}, that provide a reasonable trade-off between completeness and purity of the stellar sample; though for this initial presentation we erred on the side of purity. The selection criteria need to be satisfied in the F090W and F150W bands simultaneously. We only use the SW bands as they are shared by all three targets, allowing a common set of culling criteria.  Our initial exploration suggests that the LW photometry may improve star--galaxy separation at faint magnitude.  This important topic is being further investigated by members of our team.

Full characterization of uncertainties in resolved stellar populations studies required artificial star tests (ASTs).  ASTs consist of adding mock stars of known properties into each frame and recovering them using the same photometric procedure that is applied to the real data.  For the purposes of this paper focused on survey description, we have not included results of the ASTs. The large data volume and multi-filter nature of the data make running ASTs computationally challenging.  We will present full AST results and analyses in the upcoming \jwst\ DOLPHOT photometry paper.

\begin{figure*}[ht!]
\plotone{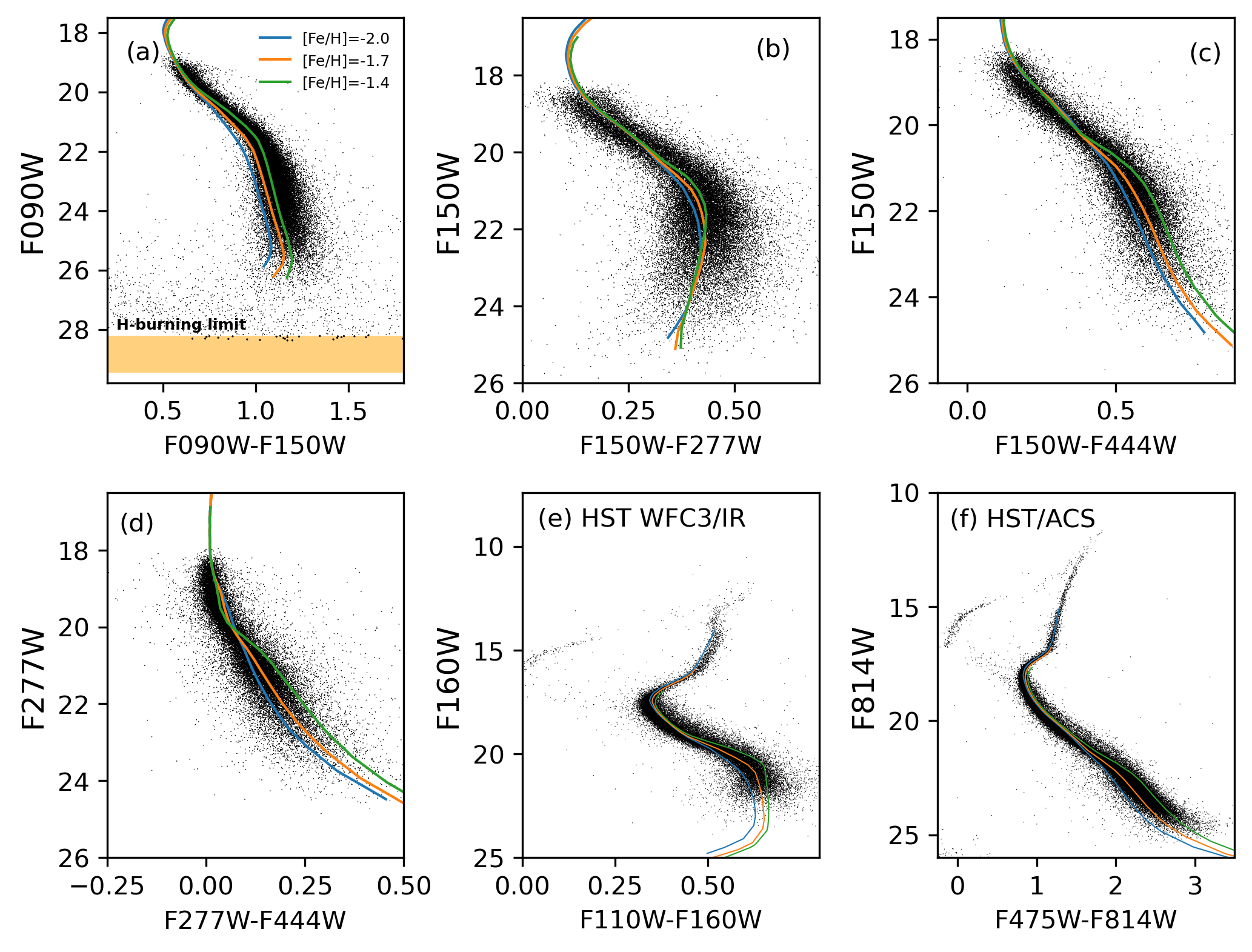}
\caption{Select NIRCam CMDs of M92, along with \hst\ WFC3/IR (\hst-GO-11664) and ACS/WFC (GO-9453, GO-10775, GO-12116, GO-16298) CMDS.  Overploted are  Solar-scaled PARSEC stellar models at fixed age (13~Gyr) over a select range of [Fe/H] values, which have been corrected to match the known $\alpha$-enhancement of M92 \citep{meszaros2020}. The SW NIRCam CMD extends from our saturation limit (F090W \, $\sim 19$) to below the lowest-mass star from the standard PARSEC models ($M=0.09\,\mathmsun$) and into the expected hydrogen burning limit regime ($0.078 \le M < 0.08\,\mathmsun$). The LW CMDs reach slightly higher limits ($M=0.12\,\mathmsun$). These are among the deepest CMDs of a GC in existence and highlight how \jwst\ will easily enable the study of prominent features for low-mass stars such as the MS-kink and lowest-mass stars.  While the stellar models are in excellent agreement with the more luminous stellar evolutionary phases (e.g., RGB, MSTO) in the \hst\ CMDs, they are $\sim0.05$~mag too blue for lower-mass stars in the SW \jwst\ filters. This offset could be due to the complex atmospheres of low-mass stars. \label{fig:m92_cmds} }
\end{figure*}

\begin{figure*}[ht!]
\plotone{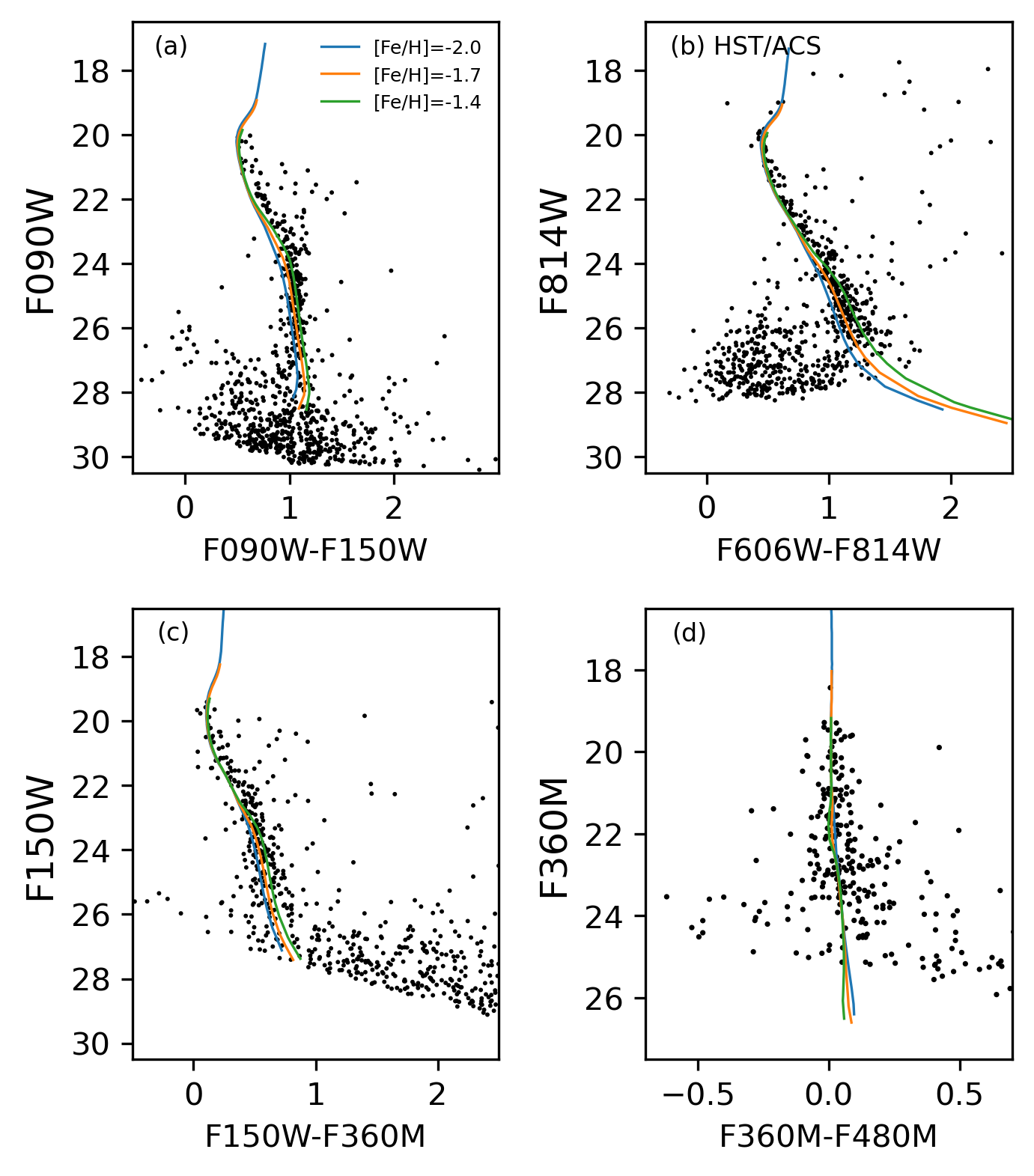}
\caption{Select NIRCam CMDs of \dracoii\ along with the deepest optical CMD which is based on \hst/ACS imaging (GO-14734; panel b).  Overploted are PARSEC stellar models at fixed age (13 Gyr) for the same [Fe/H] values selected for M92. The SW CMD extends to the lowest-mass stellar model ($M=0.09\,\mathmsun$), making it the deepest CMD of a MW satellite galaxy to date.  The exquisite depth of our data indicate how \jwst\ enables a variety of science including constraining the low-mass IMF and quantifying low-mass star features (e.g,. the MS kink and objects near the hydrogen burning limit) outside the MW.  \label{fig:draco_cmds}}
\end{figure*}

\begin{figure*}[ht!]
\plotone{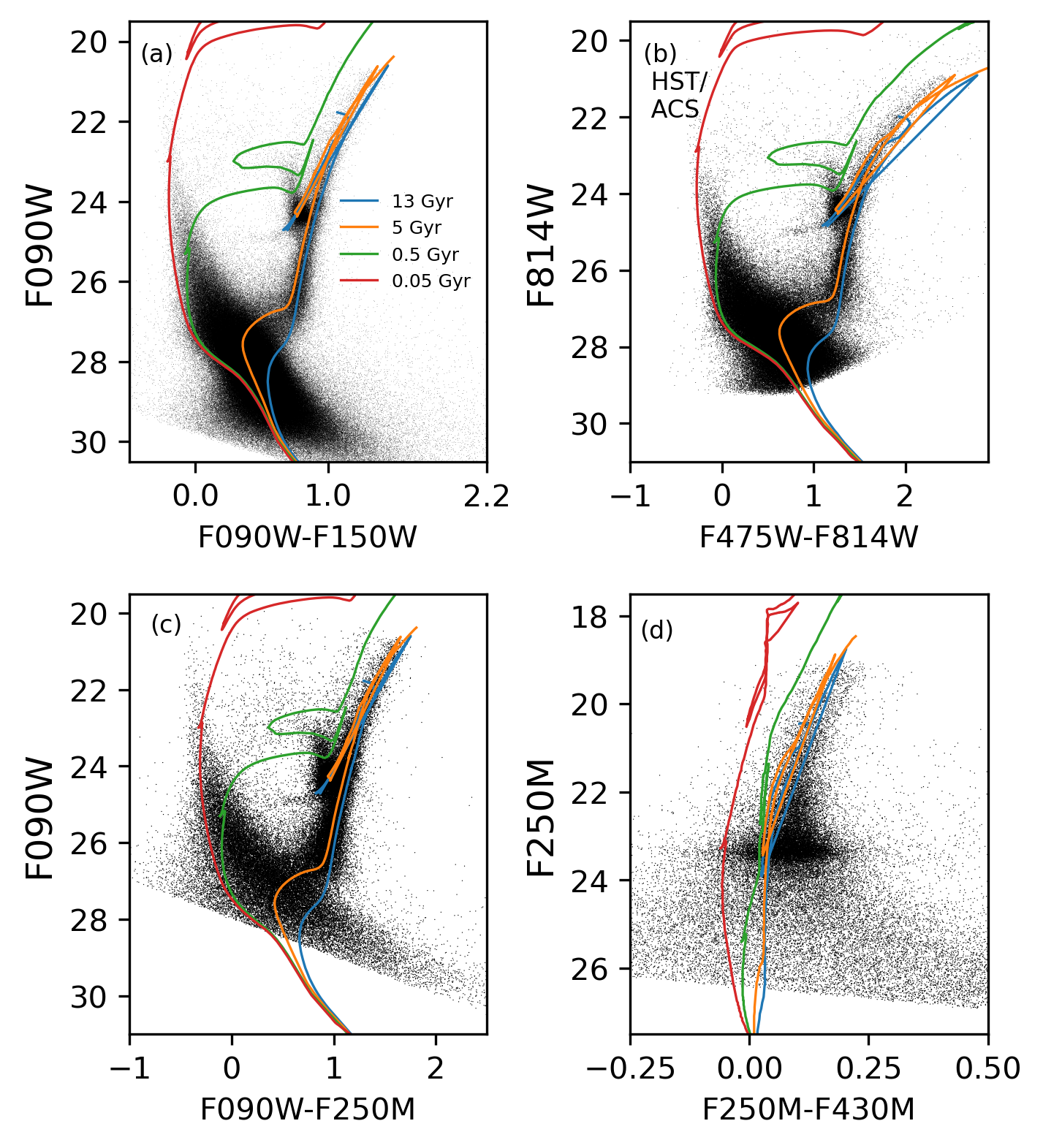}
\caption{Select NIRCam CMDs of WLM along with the deepest available \hst/ACS optical CMD from \citet{albers2019}.  Overplotted are isochrones from the PARSEC Solar-scaled stellar models at a fixed metallicity of [Fe/H]$=-1.2$~dex for a variety of indicated ages.  Panel (a) is the deepest CMD of an isolated dwarf galaxy to date, extending $\sim1.5$~mag below the oldest MSTO; it is deeper than the \hst\ CMD despite $\sim7000$s less integration time. The isochrones indicate the variety of stellar ages present in WLM.  The LW CMD shown in panel (d) extends $\sim2$~magnitudes below the red clump.  Such deep CMDs show that \jwst\ will provide for a variety of science at different cosmic epochs including exquisite lifetime SFHs, the study of evolved red stars, TRGB distances, and very young stars in galaxies outside the LG. \label{fig:wlm_cmds}}
\end{figure*}

\section{Discussion}
\label{sec:discuss}

\subsection{Color-Magnitude Diagrams}

Figure \ref{fig:all_SW_CMDs} shows the NIRCam SW CMDs for all three targets over the same magnitude range.  The juxtaposition of these CMDs illustrates the quality and diversity of science possibilities provided by our program.  In each panel, we overplot select SNRs reported by DOLPHOT. As discussed in \S \ref{sec:phot}, these SNRs are solely based on photon noise and do not account for the effects of crowding and incompletness.  We discuss SNRs further in \S \ref{sec:etc}.  

The SNRs for each target are remarkable, with SNRs ranging from 500 near the MS kink to 10 for the lowest mass stars in M92 and \dracoii.  WLM has a photon noise-based SNR of $\sim50$ at the oldest MSTO, making it the highest fidelity resolved stars observation of a distant dwarf in existence.  We now discuss the multi-band CMDs for each of our targets in more detail.

Figures \ref{fig:m92_cmds}--\ref{fig:wlm_cmds} show illustrative NIRCam CMDs in a selection of filter combinations for each ERS target.  In all cases, we apply the catalog culling parameters described in \S \ref{sec:phot} and listed in Table \ref{tab:cuts}. 

For guidance, we overplot a selection of stellar isochrones from the PARSEC v1.2S stellar libraries \citep{bressan2012,chen2014}.  These models span the full range of metallicities and ages needed to characterize our datasets. We have adjusted these isochrones to the distances and reddenings listed in Table \ref{tab:obs}.

\subsubsection{M92}
\label{sec:m92_cmds}

Figure \ref{fig:m92_cmds} shows select NIRCam CMDs for M92, along with select \hst-based CMDs for comparison. The \hst\ CMDs were reduced using DOLPHOT and the parameters recommended in \citet{williams2014}.

We overplot select PARSEC isochrones at a fixed age of 13~Gyr with varying metallicities, which we discuss below.  Though stars brighter than $F090W\sim19$ are omitted due to saturation effects (see \S \ref{sec:phot}), key CMD features for faint, low-mass stars are clearly visible.  Notably, the MS kink, which is due to opacity effects in M dwarfs, the region in which  is in panels (a) - (c) at $F090W\sim21$ and $F150W\sim20$.  The MS kink exhibits the sharpest inflection point in the F150W$-$F277W filter combination. The kink is much less pronounced in the F277W$-$F444W CMD (panel d).  This is partially due to the lower SNRs of the LW observations as well as the LW filters being far into the Rayleigh--Jeans tail of the stars' spectral energy distributions (SEDs), which makes them only weakly sensitive to stellar temperature, thus resulting in similar colors and a less obvious kink.

Our NIRCam observations have produced the deepest CMD of M92 to date.  The data extend fainter than the lowest stellar mass ($0.09\,\mathmsun$) available from the standard PARSEC stellar library in the SW filters and to slightly higher masses ($0.12\,\mathmsun$) in the LW filters.  

The faintest objects in the F090W$-$F150W CMD fall into the bright end of the expected hydrogen burning sequence.  The exact mass at which a star-like object cannot sustain hydrogen fusion has long been debated \citep[e.g.,][]{hayashi1963, kumar1963, chabrier2000}.  For this analysis, we computed custom PARSEC models with a mass resolution of 0.002 \msun\ and find that that the minimum mass for hydrogen burning is $0.078 \le M < 0.08$ \msun\ for a metallicity of [Fe/H]$=-1.7$~dex and an age of 13~Gyr. This translates to a magnitude range of $28.2 < m_{F090W} \le 29.5$~mag in M92. This depth is quite remarkable considering our observations only consistent of $\sim1050s$ in each filter.  In comparison, the faintest stars in the \hst\ WFC3/IR CMD (panel e) are a few magnitudes brighter despite $\sim1200s$ of integration time in each filter.  To date, comparably deep IR studies of metal-poor, extremely low-mass stars with \hst\ have been limited to the nearest GCs \citep[e.g., M4;][]{dieball2016, dieball2019}.  As our M92 data shows, the superior sensitivity of NIRCam will make such studies possible throughout the MW.

The PARSEC models over-plotted in Figure \ref{fig:m92_cmds} are nominally higher than the  metallicity of M92 ([Fe/H] $= -2.23$~dex) derived from high-resolution APOGEE spectroscopy \citep{meszaros2020}.  This discrepancy is because the current version of the PARSEC models are Solar-scaled, whereas M92 is highly $\alpha$-enhanced with $[\alpha$/Fe] \, $\sim0.5$~dex.  Well-established corrections can be applied to match Solar-scaled models with $\alpha$-enhanced populations \citep[e.g.,][]{salaris1993}.  For M92, the corrective factor for the PARSEC models results in values of [Fe/H] \, $\sim 0.5$~dex higher than derived from spectroscopy. PARSEC models with $\alpha$-enhancements are under construction and will mitigate the need to apply such corrections.

Overall, the PARSEC models are in reasonably good agreement with the NIRCam CMDs.  For the F150W$-$F444W and F277W$-$F444W CMDs, the models trace the locus of the data quite well.  However, for the F090W$-$F150W and F150W$-$F277W CMDs, the models are systematically too blue by $\sim0.05$~mag. The source of this offset is not due to distance, reddening, age, or metallicity as these same models are well-matched to the MSTO in the brighter \hst\ CMDs (panels e and f of Figure \ref{fig:m92_cmds}).  In general, stars above the MS kink are well-matched by the models in the LW NIRCam and \hst\ filters, limiting the offsets to only SWs.  One possible source of the offset is the presence of poorly modeled absorption features (e.g., TiO) in the atmospheres of very cool, low-mass stars.  A detailed exploration of this offset is beyond the scope of this paper, but we note that deep \jwst\ imaging of a larger set of GCs in several filters has the potential to help elucidate the exact nature of this issue. 

The NIRCam CMDs also exhibit scatter in color that is larger than photon noise and variations in age or metallicity.  The most likely explanation is the presence of multiple chemically distinct populations.  Like many luminous Galactic GCs,  M92 exhibits multiple populations with distinct abundance patterns \citep[e.g.,][]{meszaros2020}.  Other GCs with deep CMDs and similar abundance patterns show a broadening of stellar sequences at and below the MS kink.  This broadening has been attributed to the persistence of these chemically distinct sequences to the lowest stellar masses \citep[e.g.,][]{milone2012, milone2014, correnti2016, milone2017}, which is thought to be driven by oxygen variations expressed via water lines \citep[e.g.,][]{dotter2015, VandenBerg2022}.  Finally, we note that some of the scatter could be due to NIRCam zero point calibrations which will not be finalized with uncertainties until at least the end of Cycle 1 \citep{gordon2022}.

\subsubsection{Draco II}
\label{sec:draco_cmds}

Figure \ref{fig:draco_cmds} shows select NIRcam CMDs of \dracoii, along with the deepest existing optical \hst\ CMD for reference (GO-14734; PI N.\ Kallivayalil).  Like M92, the brightest stars in \dracoii\ suffer from saturation and are not included in our current photometric reduction.  We include the same PARSEC isochrones as shown in M92 (Figure \ref{fig:m92_cmds}) as \dracoii\ hosts a comparably ancient (13 Gy), metal-poor, and likely $\alpha$-enhanced stellar population \citep[e.g.,][]{longeard2018, simon2019}.  The selected isochrones provide a good fit of the MSTO in the \hst\ CMD (panel b) suggesting the adopted parameters (i.e., age, metallicity, distance, extinction) are reasonable, but as with M92, the models are slightly too blue in the NIRCam SW CMDs of \dracoii.

Our F090W$-$F150W CMD of \dracoii\ is the deepest CMD of a galaxy outside the MW and has imaged the lowest-mass stars outside the MW ($0.09\,\mathmsun$).

Previously, the deepest CMD in an external galaxy was from \citet{gennaro2018}, which used \hst\ WFC3/IR data of MW UFD Coma Berenicies to study its low-mass IMF.  From $32780s$ of integration time in each filter, their F110W and F160W photometry reached a usable low-mass limit of $0.17\,\mathmsun$, whereas our data extend to $0.09\,\mathmsun$.  The large dynamic range of stellar masses in \dracoii\ provides excellent leverage for a low-mass IMF measurement.  Tight constraints on the IMF in \dracoii\ could provide a new means of distinguishing whether faint stellar systems are dark-matter dominated dwarf galaxies or GCs \citep[e.g.,][]{willman2012, baumgardt2022}, as well as insight into star formation in extreme environments \citep[e.g.,][]{geha2013, krumholz2019}.

The width of the lower MS is in excess of photometric noise.  Metallicities derived from more luminous stars in \dracoii\ suggest a spread of $\sigma\sim0.5$~dex \citep[e.g.,][]{li2017a, longeard2018, fu2022}, which may contribute to this scatter.  Background galaxies are also a source of contamination and likely contribute to the scatter, particularly at the faintest magnitudes.  The combination of very deep imaging and the sparsity of \dracoii's stellar population mean that background galaxies are a large source of contamination.  Our preliminary investigations indicate that the multi-color NIRCam photometry may be efficient for star-galaxy separation (Warfield et al.\ in prep).

The F150$-$F360M CMD (panel c) extends to a comparably low stellar mass as the SW CMD, albeit at lower SNR. The LW CMD (panel d) is much shallower; though the primary purpose of these filters is to explore their potential as photometric metallicity indicators akin to the Calcium H \& K filters being used in the optical \citep[e.g.,][]{starkenburg2017b, longeard2018, fu2022}.

As with M92, the stellar isochrones provide a good qualitative match to the data.  The shape and magnitude of the MS kink appears to track the data well.  However, as discussed in the context of M92 the models are modestly too blue compared to the data.

\subsubsection{WLM}
\label{sec:wlm_cmds}

Figure \ref{fig:wlm_cmds} shows select NIRCam CMDs for WLM, along with the deepest optical CMD of WLM taken with \hst.  Due to its large distance, few stars in WLM are affected by saturation.  

The CMDs of WLM exhibit a wide variety of stellar sequences that span a range of ages and phases of evolution.  Examples include the young MS, the RGB and AGB, the HB, and the oldest MSTO.  The relative positions of these features are generally similar to what is known for optical CMDs with subtle changes due to the shift to IR wavelengths \citep[e.g.,][]{dalcanton2012, williams2014, gull2022}.  For example, the HB slopes to fainter values at bluer wavelengths as hot blue HB stars are less luminous at IR wavelengths.  Similarly, red stars (e.g., RGB, AGB) become more luminous as the IR wavelengths are closer to their peak temperatures compared to optical wavelengths.    

Figure \ref{fig:wlm_cmds} shows that our NIRCam SW CMD of WLM is at $\sim1$~mag deeper than the \hst/ACS CMD despite similar integration times ($\sim54200$s for NIRCam versus $\sim61400$s for ACS).  This increase owes primarily to the increased sensitivity of \jwst\ at these wavelengths.  McQuinn et al. (in prep.) is deriving the SFHs from both datasets to quantify the capabilites of \jwst\ imaging for detailed SFH determinations.  

More broadly, our CMDs of WLM are the deepest in existence for an isolated dwarf galaxy.  Prior to our program, \hst/ACS observations of Leo~A from \citet{cole2007} extended to the lowest stellar masses in an galaxy outside the immediate vicinity of the MW. The \hst\ observations of Leo~A reach nearly as deep, but the larger distance of WLM (0.5~mag farther) means that our measurements actually extend to less luminous stars on the MS.

Though not as deep as the SW data, the LW is remarkably deep for medium bands.  The F090W$-$F250M data extends below the oldest MSTO, making it the deepest medium band data available for an isolated dwarf galaxy.  The F250M$-$F430M (panel d) CMD extends well below the red clump.  This data is expected to provide an excellent means of identifying AGB stars and helping constrain their underlying physics.

The over-plotted isochrones provide a reasonable match to the data.  In this case, we have selected a single metallicity of ${\rm [Fe/H]}=-1.2$~dex matched to RGB spectroscopic abundances \citep{leaman2009}, and plotted select ages that range from 50~Myr to 13~Gyr. Visually, the PARSEC models provide a good qualitative match to the data. The level of agreement will be formally quantified in McQuinn et al. (in prep.).

\subsection{Comparisons with the Exposure Time Calculator}
\label{sec:etc}

Our photometry provides an opportunity to gauge the accuracy of the NIRCam ETC in practice.  

\begin{figure*}[ht!]
\plotone{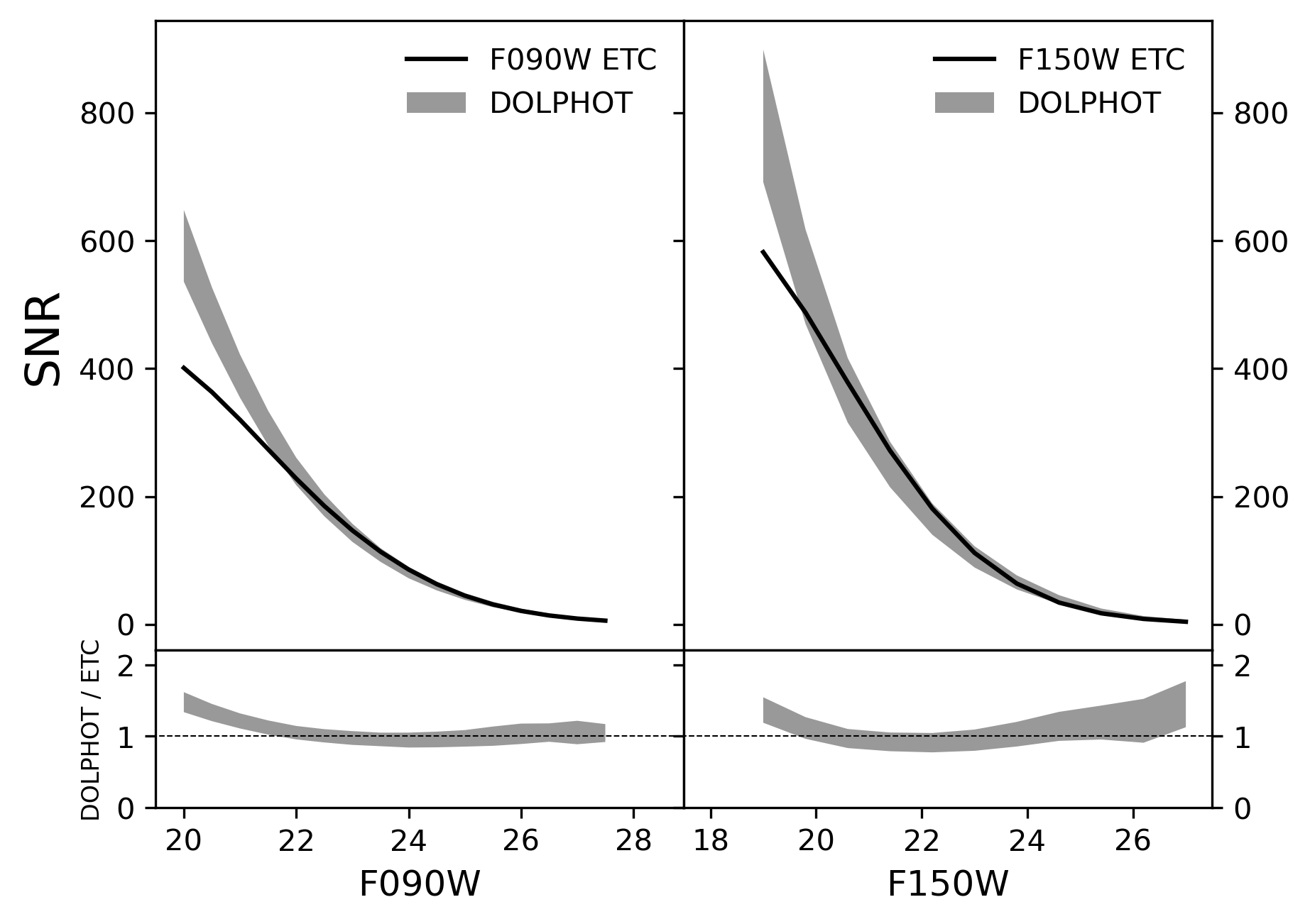}
\caption{Comparisons between the photon noise-based SNRs from the DOLPHOT F090W and F150W photometry of M92 (grey shaded region) and the  expected SNRs from v2.0 of the NIRCam ETC (black lines). Over most of the dynamic ranges in magnitude the SNRs reported by DOLPHOT and the ETC are consistent within scatter ($\sim20$\%) of the data.  Deviations at the bright end may be due to the presence of saturated pixels, while PSF shape, non-stellar objects (e.g., background galaxies), and incompleteness may contribute to a slight increase in the ratio at the faintest F150W magnitudes. \label{fig:m92_snr}}
\end{figure*}

In Figure \ref{fig:m92_snr}, we consider the SNRs for stars in M92 as reported by DOLPHOT against what is reported by the NIRCam ETC.  M92 is the best of our three targets for this exploratory exercise as it is not particularly crowded and its CMD is well-populated. The lack of crowding means that the SNRs reported by DOLPHOT are a reasonable proxy for real noise, whereas ASTs are necessary to accurately assess the noise in even moderately crowded images.  

We compute SNRs as a function of F090W and F150W by only considering stars that pass a stricter version of the culling criteria listed in \S \ref{sec:phot}.  Specifically, we require that each star has \textit{crowd} $\le0.1$, which eliminates all but the least crowded stars.  We also only consider stars with $0.2 < F090W-F150W < 1.5$, which isolates the MS in the SW filters and removes much of the contamination (e.g., background galaxies, diffraction spike artifacts) from our analysis.  Finally, we exclude the brightest stars as they may be affected by (partial) saturation.  We specifically only consider stars fainter than $F090W=20$ and $150W=19$ for their respective SNR calculations.  

From stars that pass these cuts, we compute the 50th, 16th, and 84th percentiles of the F090W and F150W SNR distributions in 0.25~mag bins over the entire magnitude ranges considered.  

For the expected SNRs, we use v2.0 of the \jwst\ ETC to compute SNR as a function of F090W and F150W. In the ETC, the detector strategies are set to match our F090W and F150W observational set up for M92 as listed in Table \ref{tab:jwst_obs} and described in \S \ref{sec:m92}. We verified that the integration times in the ETC are identical to what our program acquired.

For the ETC scene, we used a K5V star ($\Teff = 4250$~K, $\log(g) = 4.5$~dex) from the Phoenix stellar models.  Though the stellar type varies over the color and luminosity range considered, we found that reasonable changes in the choice of stellar atmosphere only affected our findings at the $\sim5$\% level. For simplicity, we adopted a single stellar atmosphere model for this calculation.

We adopted an extinction of $A_V=0.06$~mag and a MW extinction curve.  We set the background model to the central coordinates of M92 on June 20th, 2022, the date of our observations.  We computed the SNR in the F090W and F150Ws filter in 0.5~mag steps from 19 to 30~mag in each filter, renormalizing after extinction was applied.   

The result is a smooth variation in SNR as a function of magnitude.  We interpolated the results onto a finer magnitude grid for clearer comparison with the DOLPHOT results.  Interpolation errors are $<1$\%.

To compute the SNR, we used the default aperture photometry setup in the NIRCam ETC.  Specifically, this uses an aperture radius of 0.1\arcsec\ and performs background subtraction using an annulus 0.22 to 0.4\arcsec\ from the source.  For both F090W and F150W, this radius within the aperture radius range of 2-3$\times$ the PSF FWHM specified in JDOX\footnote{\url{https://jwst-docs.stsci.edu/jwst-near-infrared-camera/nircam-performance/nircam-point-spread-functions}} as recommended by the \jwst\ help desk.  We will explore more filter SNRs and variations in the ETC photometric set p in a future paper.

Figure \ref{fig:m92_snr} shows a comparison between the SNRs reported by DOLPHOT (grey shaded regions) and the NIRCam ETC (black lines) as a function of F090W and F150W magnitude.  The bottom panels show the ratios of the DOLPHOT and ETC SNRs. Both visually and quantitatively, the expected ETCs agree quite well.  For most of the magnitude ranges, the DOLPHOT and ETC SNRs agree within $\sim20$\%, which is within the bounds of our uncertainty range.  The small structures in the residuals over this range are due to finite numbers of real stars in each bin.   There are some noticeable deviations from unity at bright magnitudes for both F090W and F150W.  We believe that these may be due to saturation effects that might be mitigated by improved data quality masks and/or the use of Frame 0 data.  Both will be explored in our forthcoming DOLPHOT NIRcam module paper.  Similarly, the increased ratio at the faintest F150W magnitudes is not overly concerning as small variations in the PSF shape \citep{dolphin2000b} and the presence of non-stellar artifacts at the very bottom of the CMD can affect the photon noise-based SNRs.  The uptick could also be caused by the removal of objects that don't meet our culling criteria; formally a correct calculation requires factoring in completeness as determined by ASTs.

Overall, this comparison provides preliminary indications that v2.0 of the ETC provides reasonable SNR estimates for fairly uncrowded stars imaged with NIRCam.  Of course, in practice, many resolved stellar systems that will be targeted by NIRCam will be more affected by crowding than M92, which can lead to larger discrepancies in the expected versus recovered SNR.  For \hst, this effect is partially mitigated by the optimal SNR reported by its ETC\footnote{\url{https://etc.stsci.edu/etcstatic/users_guide/1_3_imaging.html}}.  This number reflects that expected SNR for an isolated point source recovered by PSF fitting and is generally a factor of 1.5--2 higher than the regular SNR reported by the \hst\ ETC.  Our initial analysis of M92 suggests that the baseline SNRs from the NIRCam ETC may not be off by as large a factor.  However, further exploration in a variety of images with variable crowding, stellar type, etc.  Ultimately, ASTs will aid in calculation of SNRs seen in the data over a range of stellar densities.  We will carry out such an exploration in the context of our NIRCam and NIRISS DOLPHOT photometry paper.

\section{Summary}
\label{sec:summary}

We have undertaken the \jwst\ resolved stellar populations Early Release Science program in order to establish \jwst\ as a the premier facility for resolved stellar populations early \jwst's lifetime.  In this paper, we have described the motivation, planning, implementation, execution, and present NIRCam CMDs from preliminary photometric reductions with DOLPHOT.  Some key takeaways from our survey include:

\begin{itemize}
    \item Our 27.5~hr program obtained NIRCam (primary) and NIRISS (parallel) imaging of 3 diverse targets: Milky Way globular cluster M92, satellite ultra-faint galaxy \dracoii, and more distant (0.9~Mpc) star-forming galaxy WLM.  A summary of their properties are listed in Table \ref{tab:obs} while a summary of our \jwst\ observations for each target are listed in Table \ref{tab:jwst_obs}.  These targets were selected in order to enable a variety of science and technical goals related to resolved stellar populations analysis as described in \S \ref{sec:survey}.

    \item This ERS program facilitated the development of NIRCam and NIRISS modules for DOLPHOT, a widely used stellar crowded field photometry package.  We used our ERS targets to test these modules for a variety of image properties (e.g., various filter combinations, over a large dynamic range in stellar crowding).  We describe the application of DOLPHOT to our ERS data in \S \ref{sec:phot}.  Beta versions of these DOLPHOT modules, along with theoretical PSF models for all NIRCam and NIRISS filters  are publicly available on our team website and on the DOLPHOT website. 

    \item We presented preliminary NIRCam CMDs in select SW and LW filter combinations from a first pass DOLPHOT reduction.  The CMDs are among deepest CMDs in existence for each class of object.  The F090W$-$F150W CMD of M92 touches the hydrogen burning limit ($F090W>28.2$; ($M <0.08$\,$M_{\odot}$).  The F090W$-$F150W CMD of \dracoii\ reaches the the bottom of the stellar sequence (0.09\,$M_{\odot}$) in the standard PARSEC models.  The F090W$-$F150W CMD of WLM extends $\sim$1.5~mag below the oldest MSTOs in WLM.

    \item We compare our NIRCam CMDs to select age and metallicity isochrones from the PARSEC models.  We find that the models are in generally good agreement with all \jwst\ CMDs, though we find them to be $\sim0.05$~mag systematically bluer of the lower MS in M92 and \dracoii.  We posit that this color offset may be due to the complexity of stellar atmospheres in extremely low-mass stars that is currently not well-captured in theoretical stellar atmospheres. A notable example includes the known sensitivity of color to oxygen abundance \citep[e.g.,][]{VandenBerg2022}

    \item We compare the photon-noise based SNRs for the F090W and F150W reported by DOLPHOT for stars in M92 with expectations from v2.0 of the NIRCam ETC.  We find they agree within $\sim20$\% over most of the magnitude range, with slightly larger deviations at the very bright and very faint limits.  The differences may be due to saturation effects at the bright end and selection effects and/or subtle mismatches between theoretical and observed PSFs at the faint end.  We caution that this preliminary comparison does not capture effects such as crowding, which is important in distant dwarf galaxies such as WLM.

    \item We are in the process of optimizing DOLPHOT for use with NIRCam and NIRISS.  All technical details of the DOLPHOT modules and their application to our ERS data the subject of an upcoming publication on crowded field photometry.

\end{itemize}

\begin{acknowledgments}
The authors would like to thank David W.\ Hogg for his input on the program and paper. This work is based on observations made with the NASA/ESA/CSA James Webb Space Telescope. The data were obtained from the Mikulski Archive for Space Telescopes at the Space Telescope Science Institute, which is operated by the Association of Universities for Research in Astronomy, Inc., under NASA contract NAS 5-03127 for \jwst. These observations are associated with program DD-ERS-1334.  This program also benefits from recent DOLPHOT development work based on observations made with the NASA/ESA Hubble Space Telescope obtained from the Space Telescope Science Institute, which is operated by the Association of Universities for Research in Astronomy, Inc., under NASA contract NAS 5–26555. These observations are associated with program HST-GO-15902.
\end{acknowledgments}

%

\vspace{5mm}
\facilities{JWST(NIRCAM), JWST(NIRISS)}






\bibliography{astrobib_dw_ers, citations}{}
\bibliographystyle{aasjournal}



\end{document}